\documentclass[sigconf]{acmart}

\AtBeginDocument{%
  }

\usepackage{tikz}
\usepackage{amsmath}

\usepackage{filecontents}

\usepackage{nicefrac}
\usepackage{siunitx}
\usepackage{array,framed}
\usepackage{booktabs}
\usepackage{
  color,
  float,
  epsfig,
  wrapfig,
  graphics,
  graphicx,
  subcaption
}
\usepackage{setspace}
\usepackage{latexsym,fancyhdr,url}
\usepackage{enumerate}
\usepackage[algo2e]{algorithm2e}
\usepackage{algorithmic}
\usepackage{algorithm}

\newcommand{\algorithmicbreak}{\textbf{break}}
\newcommand{\BREAK}{\STATE \algorithmicbreak}
\usepackage{graphics}
\usepackage{xparse} 
\usepackage{xspace}
\usepackage{multirow}
\usepackage{csvsimple}
\usepackage{balance}
\usepackage{mathrsfs}
\usepackage{epsfig}
\usepackage{pifont}

\usepackage{enumitem}
\usepackage{mathtools}
\usepackage{tikz}
\usepackage{pgfplots} 
\usepackage{pgfplotstable}
\pgfplotsset{compat = newest}
\usetikzlibrary{arrows}
\usetikzlibrary{patterns}
\usepackage{stackrel}
\usepackage{pstricks}
\usepackage{wrapfig}
\usepackage[utf8]{inputenc}
\usepackage{booktabs}
\usepackage{expl3}
\usepackage{collcell}
\usepackage{colortbl} 
\usepackage{xurl}
\usepackage{rotating}
\usepackage{soul}
\usetikzlibrary{
  shapes.geometric,
  arrows,
  external,
  pgfplots.groupplots,
  matrix
}

\DeclareMathAlphabet{\mathcal}{OMS}{cmsy}{m}{n}

\usepackage{amsthm}
\newtheorem{theorem}{Theorem}

\newcommand{\fullcircle}{\begin{tikzpicture}[baseline=-0.4ex]
  \draw (0,0) circle [radius=0.2em];
  \filldraw[fill=black] (0,0) circle [radius=0.2em];
\end{tikzpicture}}

\newcommand{\emptycircle}{\begin{tikzpicture}[baseline=-0.4ex]
  \draw (0,0) circle [radius=0.2em];
\end{tikzpicture}}

\newcommand{\halfcircle}{\begin{tikzpicture}[baseline=-0.4ex]
  \draw (0,0) circle [radius=0.2em];
  \filldraw[fill=black] (0,0) -- (0,0.2em) arc [start angle=90, end angle=270, radius=0.2em] -- cycle;
\end{tikzpicture}}

\DeclareGraphicsExtensions{%
    .png,.PNG,%
    .pdf,.PDF,%
    .jpg,.mps,.jpeg,.jbig2,.jb2,.JPG,.JPEG,.JBIG2,.JB2}

\newcounter{note}[section]

\newcommand{\chkmrk}{\ding{52}\xspace}
\newcommand{\xmark}{\ding{55}\xspace}%

\newcommand{\secref}[1]{\mbox{Sec.~\ref{#1}}\xspace}

\newcommand{\figref}[1]{\mbox{Fig.~\ref{#1}}\xspace}

\newcommand{\tblref}[1]{\mbox{Table~\ref{#1}}\xspace}

\newcommand{\appref}[1]{\mbox{App.~\ref{#1}}\xspace}
\newcommand{\eqnref}[1]{\mbox{Eq.~(\ref{#1})}\xspace}

\newcommand{\eqnrefstatic}[1]{\mbox{Eqn.~{#1}}}

\newcommand{\algref}[1]{\mbox{Alg.~\ref{#1}}\xspace}
\newcommand{\thmref}[1]{\mbox{Thm.~\ref{#1}}\xspace}

\def\lc{\left\lceil}   
\def\rc{\right\rceil}
\newcommand{\ceil}[1]{\ensuremath{\lc#1\rc}\xspace}

\newlength{\figureheight}

\newcommand{\Norm}[2]{\ensuremath{\left\Vert{#2}\right\Vert_{#1}}\xspace}

\newcommand{\infinityNorm}[1]{\ensuremath{\Norm{\infty}{#1}}\xspace}

\NewDocumentCommand{\genericNat}{ g }{\ensuremath{z\IfNoValueF{#1}{_{#1}}}\xspace}

\newcommand{\setSize}[1]{\ensuremath{\left|{#1}\right|}\xspace}

\newcommand{\realsPos}{\ensuremath{\mathbb{R}_{>0}}\xspace}
\newcommand{\getsr}{\ensuremath{\overset{\scriptscriptstyle\$}{\leftarrow}}\xspace}

\newcommand{\prob}[1]{\ensuremath{\mathbb{P}\left({#1}\right)}\xspace}

\NewDocumentCommand{\expv}{ o g }{\ensuremath{\mathbb{E}\IfNoValueF{#1}{_{#1}}\left({#2}\right)}\xspace}
\NewDocumentCommand{\gradient}{ g }{\ensuremath{\nabla\IfNoValueF{#1}{_{#1}}}\xspace}
\newcommand{\cset}[3]{\ensuremath{#1\{}{#2}\ensuremath{\;#1|} \ifmmode{\;}\fi {#3}\ensuremath{#1\}}\xspace}
\newcommand{\cprob}[3]{\ensuremath{\mathbb{P}#1(}{#2}\ensuremath{\;#1|} \ifmmode{\;}\fi {#3}\ensuremath{#1)}\xspace}
\newcommand{\cexpv}[3]{\ensuremath{\mathbb{E}#1(}{#2}\ensuremath{\;#1|} \ifmmode{\;}\fi {#3}\ensuremath{#1)}\xspace}

\newcommand{\genericInd}{\ensuremath{\beta}\xspace}

\newcommand{\SampleIdx}{\ensuremath{i}\xspace}
\newcommand{\TrainDataset}{\ensuremath{\mathcal{D}}\xspace}
\newcommand{\TrainDatasetSize}{\ensuremath{M}\xspace}
\newcommand{\TrainSample}[1]{\ensuremath{a_{#1}}\xspace}
\newcommand{\LossFunction}{\ensuremath{\ell}\xspace}
\NewDocumentCommand{\Data}{ g g o }{\ensuremath{x\IfNoValueF{#1}{\IfNoValueTF{#3}{_{#1}}{_{{#1},{#3}}}}\IfNoValueF{#2}{^{#2}}}\xspace}
\NewDocumentCommand{\PublishedTwin}{ g }{\ensuremath{b\IfNoValueF{#1}{_{#1}}}\xspace}
\NewDocumentCommand{\HiddenTwin}{ g }{\ensuremath{1-b\IfNoValueF{#1}{_{#1}}}\xspace}
\NewDocumentCommand{\ScoreFunction}{ g }{\ensuremath{g\IfNoValueF{#1}{^{#1}}}\xspace}
\newcommand{\TargetModel}{\ensuremath{f}\xspace}
\NewDocumentCommand{\Mark}{ g g }{\ensuremath{\delta\IfNoValueF{#1}{_{#1}}\IfNoValueF{#2}{^{#2}}}\xspace}
\NewDocumentCommand{\MarkAlt}{ g g }{\ensuremath{\delta'\IfNoValueF{#1}{_{#1}}\IfNoValueF{#2}{^{#2}}}\xspace}
\newcommand{\MarkBound}{\ensuremath{\epsilon}\xspace}
\newcommand{\PretrainedModel}{\ensuremath{h}\xspace}
\newcommand{\UtilityDistance}[2]{\ensuremath{u({#1},{#2})}\xspace}
\newcommand{\Distance}[2]{\ensuremath{d({#1},{#2})}\xspace}
\newcommand{\NullHypothesis}{\ensuremath{H_0}\xspace}
\newcommand{\AlternateHypothesis}{\ensuremath{H_1}\xspace}
\newcommand{\HypothesisProbability}{\ensuremath{\pi}\xspace}
\newcommand{\DetectionResult}{\ensuremath{\mathsf{detected}}\xspace}
\newcommand{\arrComponent}[2]{\ensuremath{\big[{#1}\big]_{#2}}\xspace}
\NewDocumentCommand{\ModelOutput}{ g g }{\ensuremath{v\IfNoValueF{#1}{_{#1}}\IfNoValueF{#2}{^{#2}}}\xspace}
\NewDocumentCommand{\Score}{ g g }{\ensuremath{s\IfNoValueF{#1}{_{#1}}\IfNoValueF{#2}{^{#2}}}\xspace}
\newcommand{\MeasurementSequence}{\ensuremath{\mathcal{S}}\xspace}
\newcommand{\PPRMartingale}{\ensuremath{\mathsf{PPRM}}\xspace}

\newcommand{\NumberQueriedPublished}{\ensuremath{Q}\xspace}
\newcommand{\MinimalPercentage}{\ensuremath{P}\xspace}
\newcommand{\NumAugment}{\ensuremath{K}\xspace}
\newcommand{\AugmentIdx}{\ensuremath{k}\xspace}
\newcommand{\AugmentIdxAlt}{\ensuremath{k'}\xspace}
\newcommand{\ClassIdx}{\ensuremath{j}\xspace}
\newcommand{\NumClasses}{\ensuremath{J}\xspace}
\newcommand{\TrainingSamples}{\ensuremath{\mathcal{X}}\xspace}
\newcommand{\PublishedInTrainDataset}{\ensuremath{\hat{\mathcal{X}}}\xspace}
\newcommand{\PublishedInTrainDatasetSize}{\ensuremath{\hat{N}}\xspace}
\NewDocumentCommand{\GroundTruthLabel}{ g g }{\ensuremath{y\IfNoValueF{#1}{_{#1}}\IfNoValueF{#2}{^{#2}}}\xspace}
\NewDocumentCommand{\Knowledge}{g}{%
  \ifthenelse{\equal{#1}{1}}{\textsc{c}\textsc{g}}%
     {\ifthenelse{\equal{#1}{2}}{\textsc{c}\bar{\textsc{g}}}%
       {\ifthenelse{\equal{#1}{3}}{\bar{\textsc{c}}{\textsc{g}}}%
         {\bar{\textsc{c}}\bar{\textsc{g}}}}}\xspace}
\newcommand{\NumQuery}{\ensuremath{\mathsf{cost}}\xspace}
\newcommand{\AccuracyDifference}{\ensuremath{\triangle\mathsf{acc}}\xspace}
\newcommand{\Accuracy}{\ensuremath{\mathsf{acc}}\xspace}
\newcommand{\DetectionSuccessRate}{\ensuremath{\mathsf{DSR}}\xspace}
\newcommand{\UBWCParameter}{\ensuremath{\tau}\xspace}
\newcommand{\GaussianNoiseParameter}{\ensuremath{\sigma}\xspace}
\newcommand{\TopK}{\ensuremath{\kappa}\xspace}
\newcommand{\TopScores}[1]{\ensuremath{\mathsf{Top{#1}}}\xspace}

\newcommand{\MemGuard}{\ensuremath{\mathsf{MemGuard}}\xspace}
\newcommand{\DifferentialPrivacy}{\ensuremath{\mathsf{DP}}\xspace}
\newcommand{\EarlyStopping}{\ensuremath{\mathsf{EarlyStop}}\xspace}
\newcommand{\AdversarialRegularization}{\ensuremath{\mathsf{AdvReg}}\xspace}
\newcommand{\TwinsDetection}{\ensuremath{\mathsf{PairDetect}}\xspace}
\newcommand{\NoTrainAugmentation}{\ensuremath{\mathsf{NoTrainAug}}\xspace}
\newcommand{\GaussianNoise}{\ensuremath{\mathsf{Gaussian}}\xspace}
\newcommand{\MarkPerturb}{\ensuremath{\mathsf{MarkPerturb}}\xspace}

\newcommand{\CosinSim}{\ensuremath{\mathsf{cosim}}\xspace}
\newcommand{\Token}[2]{\ensuremath{c_{#1}^{#2}}\xspace}
\newcommand{\SequenceLength}{\ensuremath{L}\xspace}
\newcommand{\TokenIdx}{\ensuremath{l}\xspace}
\newcommand{\TokenVocabulary}{\ensuremath{\mathcal{V}}\xspace}
\NewDocumentCommand{\Image}{ g g }{\ensuremath{\dot{x}\IfNoValueF{#1}{_{#1}}\IfNoValueF{#2}{^{#2}}}\xspace}
\NewDocumentCommand{\Caption}{ g g }{\ensuremath{\ddot{x}\IfNoValueF{#1}{_{#1}}\IfNoValueF{#2}{^{#2}}}\xspace}
\newcommand{\VisualEncoder}{\ensuremath{f'}\xspace}
\newcommand{\TextEncoder}{\ensuremath{f''}\xspace}

\newcommand{\totalSamples}{\ensuremath{N}\xspace}
\newcommand{\indicatorRV}[1]{\ensuremath{\mathbb{I}_{#1}}\xspace}

\NewDocumentCommand{\IndicatorFunction}{ g }{\ensuremath{\mathbb{I}\IfNoValueF{#1}{({#1})}}\xspace}
\NewDocumentCommand{\indicatorRVIdx}{ g }{\ensuremath{n\IfNoValueF{#1}{_{#1}}}\xspace}
\newcommand{\indicatorVal}[1]{\ensuremath{I_{#1}}\xspace}
\NewDocumentCommand{\sumIndicatorVals}{ g }{\ensuremath{\mathbb{N}\IfNoValueF{#1}{_{#1}}}\xspace}
\newcommand{\sumThreshold}{\ensuremath{T}\xspace}

\newcommand{\nullHypothesis}{\ensuremath{H_0}\xspace}
\NewDocumentCommand{\pValue}{ g }{\ensuremath{p\IfNoValueF{#1}{_{#1}}}\xspace}
\NewDocumentCommand{\FalseDetectionRate}{ g }{\ensuremath{p\IfNoValueF{#1}{_{#1}}}\xspace}
\newcommand{\BDParameter}{\ensuremath{p'}\xspace}
\newcommand{\ObservationTime}{\ensuremath{t}\xspace}
\newcommand{\UniformSample}{\ensuremath{\mathsf{Uniform}}\xspace}
\newcommand{\TotalSuccess}{\ensuremath{N'}\xspace}
\newcommand{\Success}{\ensuremath{N''}\xspace}
\newcommand{\Confidence}{\ensuremath{\alpha}\xspace}
\newcommand{\ConfidenceInterval}[2]{\ensuremath{C_{#1}({#2})}\xspace}
\newcommand{\ConfidenceIntervalLower}[2]{\ensuremath{L_{#1}({#2})}\xspace}
\newcommand{\ConfidenceIntervalUpper}[2]{\ensuremath{U_{#1}({#2})}\xspace}

\newcommand{\Carrier}[1]{\ensuremath{m_{#1}}\xspace}

\ExplSyntaxOn
\cs_new:Npn \intensity #1
   {\fp_eval:n {1.0 - (\MinIntensity + (1.0-\MinIntensity) * ((({#1}-\value{MinNumber})/(\value{MaxNumber}-\value{MinNumber}))))}
   }
\ExplSyntaxOff
\newcommand{\MinIntensity}{0.0}   
\newcounter{MinNumber}
\newcounter{MaxNumber}

\newcommand{\ApplyGradientX}[1]{\cellcolor[gray]{\intensity{#1}}}
\newcolumntype{X}{>{\collectcell\ApplyGradientX}p{1.25em}<{\endcollectcell}}

\copyrightyear{2024}
\acmYear{2024}
\setcopyright{rightsretained}
\acmConference[CCS '24]{Proceedings of the 2024 ACM SIGSAC Conference on Computer and Communications Security}{October 14--18, 2024}{Salt Lake City, UT, USA} \acmBooktitle{Proceedings of the 2024 ACM SIGSAC Conference on Computer and Communications Security (CCS '24), October 14--18, 2024, Salt Lake City, UT, USA} \acmDOI{10.1145/3658644.3690226} \acmISBN{979-8-4007-0636-3/24/10}

\makeatletter
\gdef\@copyrightpermission{
  \begin{minipage}{0.2\columnwidth}
   \href{https://creativecommons.org/licenses/by-nc-nd/4.0/}{\includegraphics[width=0.90\textwidth]{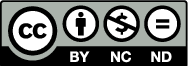}}
  \end{minipage}\hfill
  \begin{minipage}{0.8\columnwidth}
   \href{https://creativecommons.org/licenses/by-nc-nd/4.0/}{This work is licensed under a Creative Commons Attribution-NonCommercial-NoDerivs International 4.0 License.}
  \end{minipage}
  \vspace{5pt}
}
\makeatother

\begin{document}
\def\thetitle{A General Framework for Data-Use Auditing of ML Models}
\title{\thetitle}

\author{Zonghao Huang}
\orcid{0009-0006-1093-2604}
\affiliation{
  \institution{Duke University}
  \city{Durham}
  \state{NC}
  \country{USA}
}
\email{zonghao.huang@duke.edu}
\author{Neil Zhenqiang Gong}
\orcid{0000-0002-9900-9309}
\affiliation{
  \institution{Duke University}
  \city{Durham}
  \state{NC}
  \country{USA}
}
\email{neil.gong@duke.edu}
\author{Michael K. Reiter}
\orcid{0000-0001-7007-8274}
\affiliation{
  \institution{Duke University}
  \city{Durham}
  \state{NC}
  \country{USA}
}
\email{michael.reiter@duke.edu}

\renewcommand{\shortauthors}{Zonghao Huang, Neil Zhenqiang Gong, \& Michael K. Reiter}

\date{}

\begin{abstract}
  Auditing the use of data in training machine-learning (ML) models is
  an increasingly pressing challenge, as myriad ML practitioners
  routinely leverage the effort of content creators to train models
  without their permission. In this paper, we propose a general method
  to audit an ML model for the use of a data-owner's data in training,
  without prior knowledge of the ML task for which the data might be
  used. Our method leverages any existing black-box membership
  inference method, together with a sequential hypothesis test of our
  own design, to detect data use with a quantifiable, tunable
  false-detection rate.  We show the effectiveness of our proposed
  framework by applying it to audit data use in two types of ML
  models, namely image classifiers and foundation models.

\end{abstract}  



\maketitle


\section{Introduction}  \label{sec:introduction}

The advances of machine learning (ML) models hinge on the availability
of massive amounts of training data~\cite{deng2009:imagenet,
  krizhevsky2009:learning, lin2014:microsoft,
  Lle2015:TinyIV, zhang2015:character, merity2016:pointer,
  hendrycks2020:measuring}.  For
example, Contrastive Language-Image Pre-training (CLIP), developed by
OpenAI, is pretrained on 400 million of pairs of images and texts
collected from the Internet~\cite{radford2021:learning}, and large
language models like Llama 2, developed by Meta AI, are pretrained and
fine-tuned on trillions of tokens~\cite{touvron2023:llama}.  Although
the development of these large ML models has significantly contributed
to the evolution of artificial intelligence, their developers often do
not disclose the origins of their training data.  This lack of
transparency raises questions and concerns about whether appropriate
authorization to use this data to train models was obtained from their
owners.  At the same time, recent data-protection regulations, such as
the General Data Protection Regulation (GDPR) in
Europe~\cite{mantelero2013:eu}, the California Consumer Privacy Act in
the US~\cite{ccpa}, and PIPEDA privacy legislation in
Canada~\cite{cofone2020:right}, grant data owners the right to know
how their data is used.  Therefore, auditing the use of data in
ML models emerges as an urgent and important problem.

Data auditing refers to methods by which data owners can verify
whether their data was used to train an ML model.  Existing methods
include \textit{passive} data auditing and \textit{proactive} data
auditing.  Passive data auditing, commonly referred as membership
inference~\cite{shokri2017:membership, choquette2021:label,
  ye2022:enhanced, carlini2022:membership, hu2022:membership}, infers
if a data sample is a member of an ML model's training set.  However,
such passive techniques have an inherent limitation: they do not
provide any quantitative guarantee for the false-detection of their
inference results. In contrast, proactive data auditing techniques
embed marks into data before its
publication~\cite{sablayrolles2020:radioactive, li2022:untargeted,
  tang2023:did, li2023:black, guo2024:domain, wenger2024:data, wang2024:diagnosis} and
can provide detection results with false-detection
guarantees~\cite{sablayrolles2020:radioactive}. The existing proactive
data auditing methods mainly focus on \emph{dataset
auditing}~\cite{sablayrolles2020:radioactive, li2022:untargeted,
  tang2023:did, li2023:black, guo2024:domain},
where the whole training set of the ML model is contributed from one
data owner and thus the data owner has control over the whole dataset,
including, e.g., knowledge of the
labels~\cite{sablayrolles2020:radioactive, li2022:untargeted, li2023:black}.  This
limits their application in a real-world setting where the training
dataset might be collected from multiple data owners or data sources.
In addition, the existing works focus on a particular type of ML
model, e.g., image classifiers~\cite{sablayrolles2020:radioactive,
  li2022:untargeted, li2023:black, wenger2024:data}, and do not directly generalize to
other domains. Therefore, there is a need to design a general
proactive data auditing framework that requires no assumption on the
dataset curation (e.g., data labeling) and can be applied to
effectively audit data across various domains.
	
In this work, we propose such a proactive data-auditing framework.  In
a nutshell, the contribution of this framework is to turn any passive
membership-inference technique into a proactive data-auditing
technique with a quantifiable and tunable false-detection rate (i.e.,
probability of falsely detecting data use in an ML model).  Our
framework consists of a data marking algorithm and a detection
algorithm. The data marking algorithm, which the data owner applies
prior to data publication, generates \textit{two} versions of each raw
datum; each version is engineered to preserve the utility of the raw
datum from which it is generated, but otherwise the versions are
perturbed with maximally different marks. Taking the example of an
image, the marks are pixel additions to the raw image that preserve
its visual quality but maximize the difference between the two marked
versions.  Critically, this marking step is agnostic to the ML task
(including, e.g., labels) in which the data versions might be used.
The data owner then publishes only \textit{one} of the two versions,
chosen uniformly at random, and keeps the other hidden.

The key insight of our framework is that once the model is accessible
(even in only a black-box way), any ``useful'' membership-inference
technique should more strongly indicate the use of the published
version than of the unpublished version, if the model was trained
using the data-owner's published data.  If the model was \textit{not}
trained using the data-owner's published data, then the
membership-inference technique might indicate either the published or
unpublished version as more likely to have been included.  However,
because the published version was chosen uniformly, the version that
the membership test more strongly indicates was used should be equally
distributed between the two.

This insight enables us to design a sequential hypothesis test of the
null hypothesis that the ML model was \textit{not} trained using the
data-owner's data.  Using any membership-inference test, the data
owner queries the model on both the published and unpublished versions
of each datum (possibly obscured to avoid detection, as we will
discuss in \secref{sec:classifier:results:adaptive_attack}), keeping a
count of the times the published version was reported as having been
used with greater likelihood.  We derive a test to determine when the
data owner can stop and reject the null hypothesis, concluding that
the model was trained with her published data, \textit{with any
  desired false-detection rate}.

We study the performance of our proposed framework in two cases: image
classifiers and foundation models.  An image classifier is a type of
ML model used to assign labels to images based on their
content~\cite{deng2009:imagenet, simonyan2015:very, he2016:deep},
while foundation models are general-purpose, large ML
models~\cite{devlin2018:bert, brown2020:language,
  radford2021:learning, touvron2023:llama}.  In the first case, our
results on multiple visual benchmark datasets demonstrate that our
proposed framework effectively audits the use of the data-owner's data
in image classifiers across various settings. Moreover, our proposed
method outperforms the existing state-of-the-art data auditing
methods, notably Radioactive Data~\cite{sablayrolles2020:radioactive}
and Untargeted Backdoor Watermark-Clean
(UBW-C)~\cite{li2022:untargeted}.  We also investigate adaptive
attacks that the ML practitioner might use to defeat our auditing
method.  While our results show that certain adaptive attacks like
early stopping and differential privacy can degrade the detection
performance of our method, they do so at the cost of significantly
diminishing the utility of the model. For the case of foundation
models, we extended our evaluation to three types of foundation
models: a visual encoder trained by self-supervised
learning~\cite{chen2020:simple}, Llama~2~\cite{touvron2023:llama}, and
CLIP~\cite{radford2021:learning}.  Our results show that the proposed
data auditing framework achieves highly effective performance across
all of these foundation models.  Overall, our proposed framework
demonstrates high effectiveness and strong generalizability across
different types of ML models and settings.

To summarize, our contributions are as follows:
\begin{itemize}[nosep,leftmargin=1em,labelwidth=*,align=left]
    \item We propose a novel and general framework for proactive data
      auditing.  Our framework has a simple data-marking algorithm
      that is agnostic to any data labeling or ML task, and a novel
      detection algorithm that is built upon contrastive membership
      inference and a sequential hypothesis test that offers a tunable
      and quantifiable false-detection rate.
    \item We demonstrate the effectiveness of the proposed framework by
      applying it to audit the use of data in two types of ML models,
      namely image classifiers and foundation models, under various
      settings.
\end{itemize}

Source code of implementing our framework is available at
\url{https://github.com/zonghaohuang007/ML_data_auditing}.

\section{Related Work}  \label{sec:related_work}

\subsection{Data Auditing}  \label{sec:related_work:data_auditing}

Data auditing is a type of \emph{proactive} technique that a data
owner can use to audit the use of her data in a target ML
model~\cite{sablayrolles2020:radioactive, li2022:untargeted,
  tang2023:did, li2023:black, guo2024:domain, wang2024:diagnosis}.
Such methods usually include a marking algorithm that embeds marks
into data, and a detection algorithm that tests for the use of that
data in training a model.  Radioactive
Data~\cite{sablayrolles2020:radioactive} is a state-of-the-art method
for auditing an image classifier, which we consider as one of our
baselines in \secref{sec:classifier:setup}.  In the marking step,
Radioactive Data randomly samples \emph{class-specific} marks and
embeds them into a subset of the training dataset.  In the detection
step, it detects if the parameters of the final layer of the target
image classifier are correlated with the selected marks, by a
hypothesis test whose returned p-value is its false-detection rate.
However, Radioactive Data assumes that the data owner has full control
over the training set, including that, e.g., she knows the labels of
the dataset and can train a surrogate model (i.e., a model similar to
the target model) used to craft marked images. In contrast, our
  work relaxes the requirement for one data owner to control the
  entire training dataset.  Another marking-based technique that,
  like ours, relaxes this requirement for classifiers is that of
  Wenger, et al.~\cite{wenger2024:data}.  However, unlike ours, this
  technique requires most of marked data contributed by a data owner
  to be assigned the same label by the ML practitioner; does not
  provide a rigorous guarantee on the false-detection rate; and to
  achieve good detection performance in their reported experiments on
  image classifiers, needed marks that were sufficiently visible to
  diminish image quality.
  
Another line of works on image dataset
  auditing~\cite{li2022:untargeted, tang2023:did, li2023:black} is
  based on backdoor attacks~\cite{gu2019:badnets, saha2020:hidden} or other methods
  (e.g.,~\cite{guo2024:domain}) to enable a data-owner to modify her
  data and then detect its use to train an ML model by eliciting
  predictable classification results from the model (e.g., predictable
  misclassifications of poisoned images for backdoor-based methods).
  Their detection algorithms are also formulated by a hypothesis test,
  but they do not provide rigorous guarantees on their false-detection
  rates. Moreover, these methods again require the data owner's full
  control over the training set, in contrast to our method.  In
  \secref{sec:classifier:setup}, we consider one backdoor-based
  auditing method, namely Untargeted Backdoor Watermark-Clean
  (UBW-C)~\cite{li2022:untargeted}, as one of our baselines.

To our knowledge, all existing data auditing methods focus on a
particular type of ML model, e.g., image
classifiers~\cite{sablayrolles2020:radioactive, li2022:untargeted,
  tang2023:did, li2023:black, wenger2024:data, guo2024:domain},
language models~\cite{wei2024:proving}, or text-to-image diffusion
models~\cite{wang2024:diagnosis}.  So, their proposed techniques do
not directly generalize to other domains.  In contrast, the marking
algorithm in our proposed framework does not rely on any prior
knowledge of the ML task (e.g., labels assigned by the ML
practitioner), and our framework can be used to effectively audit data
across various domains.

\subsection{Membership Inference}  \label{sec:related_work:membership_inference}

Membership inference (MI) is a type of confidentiality attack in
machine learning, which aims to infer if a particular data
sample~\cite{shokri2017:membership, choquette2021:label,
  ye2022:enhanced, carlini2022:membership, hu2022:membership} or any
data associated with a specific
user~\cite{song2019:auditing,miao2019:audio,chen2023:face} has been
used to train a target ML model.  The existing MI methods can be
classified into shadow model-based
attacks~\cite{shokri2017:membership, long2020:pragmatic} and
metric-based attacks~\cite{yeom2018:privacy, salem2019:ml,
  sablayrolles2019:white, song2021:systematic}.  Shadow model-based
attacks leverage shadow models (i.e., models trained on datasets that
are similar to the training dataset of the target model) to imitate
the target model and so incur high costs to train them.  In contrast,
metric-based attacks leverage metrics that are simple to compute
(e.g., entropy of the confidence vector output by the target
classifier~\cite{salem2019:ml, song2021:systematic}) while achieving
comparable inference performance~\cite{yeom2018:privacy, salem2019:ml,
  sablayrolles2019:white}.  MI has been explored for various model
types, e.g., image classifiers~\cite{yeom2018:privacy, salem2019:ml,
  sablayrolles2019:white, song2021:systematic}, visual encoders
trained by self-supervised learning~\cite{liu2021:encodermi}, language
models~\cite{pan2020:privacy}, reinforcement
learning~\cite{du2024:orl}, and facial recognition
models~\cite{chen2023:face}.

MI can be used as a \emph{passive} data auditing method that a data
owner can use to infer if her data is used in an ML model. However,
such a passive method does not provide any quantitative guarantee for
its inference results.  Our proposed framework uses metric-based MI to
design the score function in the detection algorithm that provides a
quantifiable, tunable guarantee on false detection.

\subsection{Data Watermarking}  \label{sec:related_work:data_marking}

Data watermarking is a technique used to track digital data by
embedding a watermark that contains identifying information of the
data owner. A classical example of image watermarking is zero-bit
watermarking~\cite{cayre2005:watermarking} that embeds information
into the Fourier transform of the image. However, this type of
traditional watermarking is not robust to data transformation.
Recently, there have been research efforts on training deep neural
networks (DNNs) to embed and recover watermarks that are robust to
data transformation~\cite{baluja2017:hiding, zhu2018:hidden,
  luo2020:distortion, tancik2020:stegastamp}.  DNN-based data
watermarking is widely applied to attribute AI-generated
content~\cite{yu2021:deepfake, fernandez2023:stable}.

Data watermarking can be used to audit data use to train a generative
model~\cite{yu2021:deepfake}, since the watermark embedded in the
training images could be transferred to the images generated from the
model. However, this technique cannot be directly applied to other
types of ML models, e.g., an image classifier.  In contrast, instead
of recovering the embedded marks from the ML model, our proposed
auditing method detects the use of published data by analyzing the
outputs of the ML model on the published data and the hidden data.

\section{Problem Formulation} \label{sec:problem}

We consider two parties: a \textit{data owner} and a \textit{machine
  learning practitioner}.  The data owner holds a set $\{\Data{1},
\Data{2}, \dots, \Data{\totalSamples}\}$ of data that will be
published online, e.g., posted on social media to attract attention.
The ML practitioner aims to train a machine learning model
\TargetModel of good utility on a set of training data $\TrainDataset
= \{\TrainSample{\SampleIdx}\}_{\SampleIdx=1}^{\TrainDatasetSize}$ of
size $\TrainDatasetSize$ by solving:
\begin{equation} \label{eq:training}
    \min_{\TargetModel} \frac{1}{\TrainDatasetSize}\sum_{\SampleIdx=1}^{\TrainDatasetSize} \LossFunction(\TargetModel, \TrainSample{\SampleIdx}),
\end{equation}
where \LossFunction is a loss function used to measure the performance
of the ML model on the training samples.  The definition of the loss
function depends on the machine learning task. For example, the loss
function in image classification is the cross-entropy loss~\cite{mao2023:cross}.

\subsection{Threat Model} \label{sec:problem:threat_model}

The ML practitioner wants to assemble a training dataset \TrainDataset
that can be used to train a useful ML model. He does so by collecting
the data published online from multiple data owners, \emph{without}
their authorization. As such, a data owner's data constitutes a subset
of the ML practitioner's collected dataset (i.e., some portion of
$\{\Data{1}, \Data{2}, \dots, \Data{\totalSamples}\}$ or its published
version is contained in \TrainDataset). The ML practitioner
preprocesses the collected data (e.g., labeling it, if needed), trains
an ML model on the preprocessed data using a learning algorithm
specified for his ML task (e.g., supervised learning for image
classification), and deploys it to provide service to consumers.

The data owner wants to detect the ML practitioner's use of her data.
To do so, the data owner needs to apply a method to audit the ML
practitioner's ML model such that if the ML model uses her published
data, then she will detect this fact from the deployed model. We allow
the data owner only black-box access to the deployed ML model.  In
other words, she does not necessarily know the architecture and
parameters of the ML model, but can obtain the outputs of the ML model
by providing her queries, e.g., predictions or vectors of confidence
scores output by an image classifier given her images as inputs.

\subsection{Design Goals} \label{sec:problem:design_goal}
In this work, we aim to design a \emph{data auditing} framework for a
data owner, which she can apply to detect the ML practitioner's use of
her data. We have the following design goals for the proposed data
auditing framework:
\begin{itemize}[nosep,leftmargin=1em,labelwidth=*,align=left]
    \item \textbf{Effectiveness}: The main goal of the proposed data
      auditing framework is to detect the unauthorized use of data in
      ML model training. When the published data is used, the proposed
      method should successfully detect the use of the owner's data.
      More specifically, the detection success rate (i.e., the
      probability of successfully detecting the data use) should grow
      with the amount of the owner's data that the ML practitioner
      uses in training, and should approach $100\%$ if most of her
      data is used.
    \item \textbf{Quantifiable false-detection rate}: When the ML
      practitioner does \textit{not} use the owner's data, then
      detection should occur with only a quantifiable probability
      (e.g., $\le 5\%$).  Such false-detection rate guarantees that if
      the ML practitioner does not use the data owner's data, then the
      risk of falsely accusing him is small and quantifiable.
    \item \textbf{Generality}: Once the data owner publishes her data
      online, the ML practitioner might collect them, label them if
      needed, and use them in the ML-model training for his designed
      ML task. The generality goal is that the algorithm applied prior
      to data publication (i.e., the data-marking algorithm,
      introduced in \secref{sec:proposed_framework:marking}) should be
      agnostic to the data labeling and the ML task, and that the
      proposed data auditing framework can be applied to effectively
      audit data in any type of ML model (e.g., image classifier or
      language model).
    \item \textbf{Robustness}: Once the ML practitioner realizes that
      the data auditing method is applied, he would presumably deploy
      countermeasures/adaptive attacks to defeat the data auditing
      method without sacrificing the utility of the trained ML model
      significantly. The robustness goal requires that the proposed
      framework is still effective to detect the unauthorized use of
      data in model training even when utility-preserving
      countermeasures/adaptive attacks have been applied by the ML
      practitioner.
\end{itemize}

\section{The Proposed Framework} \label{sec:proposed_framework}

In this section, we propose a framework used to detect if an ML model
has been trained on the data owner's data.  In our framework, the data
owner does not publish her data $\{\Data{1}, \Data{2}, \dots,
\Data{\totalSamples}\}$ directly.  Instead, she creates two different
marked versions of each \Data{\SampleIdx}, namely \Data{\SampleIdx}{0}
and \Data{\SampleIdx}{1}; uniformly randomly chooses a bit
$\PublishedTwin{\SampleIdx} \getsr \{0,1\}$; and publishes
\Data{\SampleIdx}{\PublishedTwin{\SampleIdx}} while keeping
\Data{\SampleIdx}{\HiddenTwin{\SampleIdx}} private.  If the ML
practitioner's ML model \TargetModel is \emph{not} trained on the
published data, it will behave equally when provided the published
data and the unpublished data as input (e.g., for classification).
Otherwise, its behavior will be biased towards the published data due
to their memorization in training~\cite{chatterjee2018:learning,
  song2017:machine}.

Formally, let \ScoreFunction{\TargetModel} denote a score function
\ScoreFunction with oracle (i.e., black-box) access to ML model
\TargetModel and that is designed for black-box membership
inference~\cite{song2021:systematic,choquette2021:label,
  liu2021:encodermi}, so that its output (a real number) indicates the
likelihood that its input was a training sample for \TargetModel.  If
the ML model \TargetModel is \emph{not} trained on the published data
\Data{\SampleIdx}{\PublishedTwin{\SampleIdx}}, then the probability of
the event
$\ScoreFunction{\TargetModel}(\Data{\SampleIdx}{\PublishedTwin{\SampleIdx}})
>
\ScoreFunction{\TargetModel}(\Data{\SampleIdx}{\HiddenTwin{\SampleIdx}})$
will be $\frac{1}{2}$; otherwise, the probability will be larger than
$\frac{1}{2}$.  The probability $\frac{1}{2}$ is due to the uniformly
random sampling of \PublishedTwin{\SampleIdx}. As such, we can detect
if an ML model is trained on a dataset containing a subset of
published data by observing the different performance of the ML model
on the published data and the unpublished data.  Since we compare the
membership inference scores (likelihoods) of published data and
unpublished data, we refer to this technique as using
\textit{contrastive} membership inference.  When the published data is
used in training, a ``useful'' membership inference will give a higher
score to published data than to unpublished data, even though both
scores might be high enough to predict them as ``members''
independently.  More details on how to generate the published data and
the unpublished data and how to measure the bias in the ML model will
be discussed later (see \secref{sec:proposed_framework:marking} and
\secref{sec:proposed_framework:detection}, respectively).

Generally, our framework includes a marking algorithm and a detection
algorithm. The marking algorithm is applied in the marking step before
the data publication, while the detection algorithm is applied in the
detection step after the ML model deployment.

\subsection{Data Marking} \label{sec:proposed_framework:marking}
The marking algorithm, applied in the marking step, is used to
generate a pair of published data and unpublished data. Formally, the marking
algorithm takes as input a raw datum \Data{\SampleIdx}, and outputs
its published version \Data{\SampleIdx}{\PublishedTwin{\SampleIdx}}
and its unpublished version
\Data{\SampleIdx}{\HiddenTwin{\SampleIdx}}. The marking algorithm
includes a \emph{marked data generation} step and a \emph{random
sampling} step, and its pseudocode is presented in~\algref{alg:marking} 
in \appref{app:algorithm}. The marked
data generation step creates a pair $(\Data{\SampleIdx}{0},
\Data{\SampleIdx}{1})$, both crafted from the raw datum
\Data{\SampleIdx}.  Taking the example where \Data{\SampleIdx} is an
image, we set $\Data{\SampleIdx}{0} \gets \Data{\SampleIdx} +
\Mark{\SampleIdx}$ and $\Data{\SampleIdx}{1} \gets \Data{\SampleIdx} -
\Mark{\SampleIdx}$ where $\Mark{\SampleIdx}$ is the added mark.  The
random sampling step selects $\PublishedTwin{\SampleIdx} \getsr \{0,
1\}$ and publishes \Data{\SampleIdx}{\PublishedTwin{\SampleIdx}},
keeping \Data{\SampleIdx}{\HiddenTwin{\SampleIdx}} secret.

\paragraph{Basic requirements} We have the following requirements
for the generated \Data{\SampleIdx}{0} and \Data{\SampleIdx}{1}:
\emph{utility preservation} and \emph{distinction}.
\begin{itemize}[nosep,leftmargin=1em,labelwidth=*,align=left]
\item \emph{Utility preservation}: \Data{\SampleIdx}{0} and
  \Data{\SampleIdx}{1} should provide the same utility as
  \Data{\SampleIdx} to the data owner, for the purposes for which the
  data owner wishes to publish \Data{\SampleIdx} (e.g., to attract
  attention on social media).  Formally, given a well-defined distance
  function \UtilityDistance{\cdot}{\cdot} measuring the utility
  difference, utility preservation requires that
  $\UtilityDistance{\Data{\SampleIdx}{0}}{\Data{\SampleIdx}} \leq
  \MarkBound$ and
  $\UtilityDistance{\Data{\SampleIdx}{1}}{\Data{\SampleIdx}} \leq
  \MarkBound$, where \MarkBound is a small scalar. Taking the example
  of images, the utility distance function could be defined as the
  infinity norm of the difference in the pixel values, i.e.,
  $\UtilityDistance{\Data{\SampleIdx}{0}}{\Data{\SampleIdx}} =
  \infinityNorm{\Data{\SampleIdx}{0} - \Data{\SampleIdx}}$ and
  $\UtilityDistance{\Data{\SampleIdx}{1}}{\Data{\SampleIdx}} =
  \infinityNorm{\Data{\SampleIdx}{1} - \Data{\SampleIdx}}$.

\item \emph{Distinction}: \Data{\SampleIdx}{0} and
  \Data{\SampleIdx}{1} should be different enough such that
  contrastive membership inference can distinguish between a model
  trained on one but not the other.  Formally, given a well-defined
  distance function $\Distance{\cdot}{\cdot}$, distinction requires
  that $\Distance{\Data{\SampleIdx}{0}}{\Data{\SampleIdx}{1}}$ is
  maximized. Continuing with the example of images, we could define
  $\Distance{\Data{\SampleIdx}{0}}{\Data{\SampleIdx}{1}} =
  \Norm{2}{\PretrainedModel(\Data{\SampleIdx}{0}) -
    \PretrainedModel(\Data{\SampleIdx}{1})}$, where \PretrainedModel
  is an image feature extractor, e.g., ResNet18~\cite{he2016:deep}
  pretrained on ImageNet~\cite{deng2009:imagenet}.\footnote{An image
  feature extractor is not necessary but helpful to craft marked
  images.  Our proposed method can audit image data effectively even
  if no image feature extractor is used in marked data generation, as
  shown in \appref{app:hyperparameters}.}
\end{itemize}

There exists a tension between utility preservation and
distinction. Specifically, when the marked data preserves more utility
of the raw data, i.e., by using a smaller $\MarkBound$, the difference
between the two marked versions is smaller and thus it is harder for
contrastive membership inference to distinguish between a model
trained on one but not the other. In experiments in
\secref{sec:classifier} and \secref{sec:foundation_model}, we show
that we can balance utility preservation and distinction well, by
setting an appropriate $\MarkBound$.  We also analyze and discuss this
tension in \secref{sec:classifier:results}.

\paragraph{Marked data generation} To craft a pair of marked data
that satisfy the basic requirements, we formulate an optimization
problem:
\begin{subequations} \label{eq:twins_data}
  \begin{align}
    \max_{\Data{\SampleIdx}{0}, \Data{\SampleIdx}{1}} & \quad \Distance{\Data{\SampleIdx}{0}}{\Data{\SampleIdx}{1}} \\
    \text{subject to:} & \quad \UtilityDistance{\Data{\SampleIdx}{0}}{\Data{\SampleIdx}} \leq \MarkBound \quad \text{and} \quad \UtilityDistance{\Data{\SampleIdx}{1}}{\Data{\SampleIdx}} \leq \MarkBound
  \end{align}
\end{subequations}
The definitions of $\UtilityDistance{\cdot}{\cdot}$ and
$\Distance{\cdot}{\cdot}$, and how to solve \eqnref{eq:twins_data}
depend on the type of data. We will instantiate them in our
experiments in \secref{sec:classifier} and
\secref{sec:foundation_model}.

\paragraph{Random sampling} After crafting the marked data
$(\Data{\SampleIdx}{0}, \Data{\SampleIdx}{1})$, the data owner selects
$\PublishedTwin{\SampleIdx} \getsr \{0, 1\}$. Then she publishes
\Data{\SampleIdx}{\PublishedTwin{\SampleIdx}}, e.g., on social media.
She keeps \Data{\SampleIdx}{\HiddenTwin{\SampleIdx}} secret to use in
the detection step, as discussed in
\secref{sec:proposed_framework:detection}.

\subsection{Data-Use Detection} \label{sec:proposed_framework:detection}

The detection algorithm, applied in the detection step, is used to
detect if a target ML model is trained on a dataset containing the
published data. Formally, given oracle (black-box) access to an ML
model \TargetModel, the detection algorithm takes as input the data
owner's published data
$\{\Data{\SampleIdx}{\PublishedTwin{\SampleIdx}}\}_{\SampleIdx =
  1}^{\totalSamples}$ and her unpublished data
$\{\Data{\SampleIdx}{\HiddenTwin{\SampleIdx}}\}_{\SampleIdx =
  1}^{\totalSamples}$, and outputs a Boolean value. It detects the
difference between the outputs of the target ML model on the published
data and unpublished data.  Specifically, for a given $\SampleIdx$,
the data owner measures if
\begin{equation}
\ScoreFunction{\TargetModel}(\Data{\SampleIdx}{\PublishedTwin{\SampleIdx}})
>
\ScoreFunction{\TargetModel}(\Data{\SampleIdx}{\HiddenTwin{\SampleIdx}}),
\label{eq:test}
\end{equation}
where the score function $\ScoreFunction$ is a black-box membership
inference algorithm that measures the likelihood of the input being
used as a member of the training set of the target ML model
$\TargetModel$.  A higher score returned by the score function
indicates a higher likelihood, and thus the choice of the score
function depends on the type of the target ML model; we will give
examples in \secref{sec:classifier} and \secref{sec:foundation_model}.
Under the null hypothesis \NullHypothesis that the ML model
\TargetModel was not trained on the data owner's published data,
\eqnref{eq:test} holds with probability $\HypothesisProbability =
\frac{1}{2}$, where the probability is with respect to the choice of
$\PublishedTwin{\SampleIdx}$.  If it was trained on the data owner's
published data (the alternative hypothesis \AlternateHypothesis),
however, then it is reasonable to expect that \eqnref{eq:test} holds
with probability $\HypothesisProbability > \frac{1}{2}$, since the ML
model memorizes the published data. As such, the detection problem can
be formulated to test the following hypothesis:
\begin{itemize}[nosep,leftmargin=1em,labelwidth=*,align=left]
    \item Null hypothesis \NullHypothesis: $\HypothesisProbability =
      \frac{1}{2}$.
    \item Alternate hypothesis \AlternateHypothesis:
      $\HypothesisProbability > \frac{1}{2}$.
\end{itemize}
We denote the sum of successful measurements in the population as
$\TotalSuccess$, i.e., $\TotalSuccess =
\sum_{\SampleIdx=1}^{\totalSamples}
\IndicatorFunction{\ScoreFunction{\TargetModel}(\Data{\SampleIdx}{\PublishedTwin{\SampleIdx}})>
  \ScoreFunction{\TargetModel}(\Data{\SampleIdx}{\HiddenTwin{\SampleIdx}})}$
where $\IndicatorFunction$ is the indicator function returning $1$ if
the input statement is true or returning $0$ if the input statement is
false.  Under \NullHypothesis, $\TotalSuccess$ follows a binomial
distribution with parameters $\totalSamples$ and
$\BDParameter=\frac{1}{2}$.  As such, the data owner can reject
\NullHypothesis or not based on the measured $\TotalSuccess$
using a binomial test. In other words, the data owner detects if
the ML model is trained on her published data according to
$\TotalSuccess$.

\subsubsection{Estimate $\TotalSuccess$ by Sampling Sequentially WoR}
\label{sec:proposed_framework:detection:wor}

Measuring $\TotalSuccess$ exactly requires querying all the published
data and hidden data to the ML model, e.g., via its API interface.
When \totalSamples is large, this would be highly costly and time
consuming.  To address this, we apply a sequential method: at each
time step, the data owner samples an $\SampleIdx$ uniformly at random
without replacement (WoR) and estimates $\TotalSuccess$ based on the
currently obtained measurements.  The classical sequential hypothesis
testing method, namely the sequential probability ratio
test~\cite{wald1992:sequential}, requires knowing the probability
\HypothesisProbability in the alternate hypothesis
\AlternateHypothesis and so does not fit our problem.

\paragraph{Sampling WoR problem} There are \totalSamples fixed but
unknown objects in the finite population $\{\indicatorVal{1}, \ldots,
\indicatorVal{\totalSamples}\}$, where each
$\indicatorVal{\SampleIdx}$ takes on a value in $\{0, 1\}$,
specifically $\indicatorVal{\SampleIdx}=
\IndicatorFunction{\ScoreFunction{\TargetModel}(\Data{\SampleIdx}{\PublishedTwin{\SampleIdx}})
  >
  \ScoreFunction{\TargetModel}(\Data{\SampleIdx}{\HiddenTwin{\SampleIdx}})}$.
The data owner observes one object per time step by sampling it
uniformly at random WoR from the population, so that:
\begin{equation*}
  \indicatorRV{\ObservationTime} \mid \{\indicatorRV{1}, \ldots, \indicatorRV{\ObservationTime-1}\} \sim \UniformSample(\{\indicatorVal{1}, \ldots,
  \indicatorVal{\totalSamples}\} \setminus  \{\indicatorRV{1}, \ldots, \indicatorRV{\ObservationTime-1}\}),
\end{equation*}
where $\indicatorRV{\ObservationTime}$ denotes the object sampled at
time $\ObservationTime\in\{1, 2, \dots, \totalSamples\}$. As such, the
variable $\sumIndicatorVals{\ObservationTime} =
\sum_{\indicatorRVIdx=1}^{\ObservationTime}
\indicatorRV{\indicatorRVIdx}$ at time \ObservationTime
($\ObservationTime \leq \totalSamples$) follows a hypergeometric
distribution:
\begin{equation*}
  \prob{\sumIndicatorVals{\ObservationTime} = \Success} =
   {\TotalSuccess \choose \Success}{\totalSamples - \TotalSuccess \choose \ObservationTime - \Success} 
   \bigg/ {\totalSamples \choose \ObservationTime},
\end{equation*} 
where $\Success \in \{0, 1, \dots, \min(\TotalSuccess,
\ObservationTime)\}$ is the number of ones from the obtained
observations at $\ObservationTime$, and ${\TotalSuccess \choose
  \Success}$ denotes $\TotalSuccess$ choose $\Success$.

\paragraph{Estimate $\TotalSuccess$ by prior-posterior-ratio
  martingale (PPRM)~\cite{waudby2020:confidence}} In the above problem
of sampling WoR from a finite population, the data owner can use a
prior-posterior-ratio martingale (PPRM)~\cite{waudby2020:confidence}
to obtain a confidence interval
$\ConfidenceInterval{\ObservationTime}{\Confidence} =
[\ConfidenceIntervalLower{\ObservationTime}{\Confidence},
  \ConfidenceIntervalUpper{\ObservationTime}{\Confidence}]$ for
\TotalSuccess at the time $\ObservationTime$, which is a function of
the confidence level $\Confidence$, e.g., $\Confidence = 0.05$.  Such
a sequence of confidence intervals
$\{\ConfidenceInterval{\ObservationTime}{\Confidence}\}_{\ObservationTime
  \in \{1, 2, \dots, \totalSamples\}}$ has the following
guarantee~\cite{waudby2020:confidence}:
\begin{equation*}
  \prob{\exists \ObservationTime \in \{1, 2, \dots, \totalSamples\}: \TotalSuccess \notin \ConfidenceInterval{\ObservationTime}{\Confidence}} \leq \Confidence.
\end{equation*}
In words, the probability that there exists a confidence interval
where \TotalSuccess is excluded is no larger than \Confidence.

\subsubsection{Detection Algorithm with Quantifiable False-Detection Rate}
We present the pseudocode of our detection algorithm in
\algref{alg:detection} in \appref{app:algorithm}.  At each time step,
the data owner samples an $\SampleIdx \in \{1, \ldots,
\totalSamples\}$ uniformly at random WoR and estimates $\TotalSuccess$
based on the currently obtained measurements using a
prior-posterior-ratio martingale (PPRM)~\cite{waudby2020:confidence}
that takes as inputs the sequence of measurements so far, the size of
the population $\totalSamples$, and the confidence level
$\Confidence$.  It returns a confidence interval for $\TotalSuccess$.
If the interval (i.e., its lower bound) is equal to or larger than a
preselected threshold $\sumThreshold$, the data owner stops sampling
and rejects the null hypothesis; otherwise, she continues the
sampling.

Since the detection algorithm rejects the null hypothesis as long as
the lower bound of a confidence interval is equal to or larger than a
preselected threshold $\sumThreshold$, the false-detection probability
is $\cprob{\big}{\exists \ObservationTime \in \{1, 2, \dots,
  \totalSamples\}:\ConfidenceIntervalLower{\ObservationTime}{\Confidence}
  \geq \sumThreshold}{\nullHypothesis}$.  We prove the following
theorem in \appref{app:derivation}.
\begin{theorem}[False detection rate] \label{theorem:fdr}
  For $\sumThreshold \in \{\ceil{\frac{\totalSamples}{2}}, \ldots,
  \totalSamples\}$ and $\Confidence < \FalseDetectionRate$ such that
  $\bigg(\frac{\exp\big(\frac{2\sumThreshold}{\totalSamples} -
    1\big)}{\big(\frac{2\sumThreshold}{\totalSamples}\big)^{\frac{2\sumThreshold}{\totalSamples}}}\bigg)^{\frac{\totalSamples}{2}}
  \leq \FalseDetectionRate - \Confidence$, our data-use detection algorithm has a
  false-detection rate less than \FalseDetectionRate. In other words:
  \begin{equation*}
    \cprob{\big}{\exists \ObservationTime \in \{1, 2, \dots, \totalSamples\}: \ConfidenceIntervalLower{\ObservationTime}{\Confidence} \geq \sumThreshold}{\nullHypothesis} < \FalseDetectionRate.
  \end{equation*}
\end{theorem}

\section{Auditing Image Classifiers}  \label{sec:classifier}

In this section, we apply our data-use auditing method to detect
unauthorized use of data to train an image classifier.  Image
classification (e.g.,~\cite{deng2009:imagenet, simonyan2015:very,
  he2016:deep}) is a fundamental computer-vision task in which the ML
practitioner trains a model (i.e., image classifier) on training data
partitioned into $\NumClasses$ classes.  For a newly given image, the
ML model predicts a class label for it or, more generally, a vector of
$\NumClasses$ dimensions. The output vector could be a vector of
confidence scores whose \ClassIdx-th component represents the
probability of the input being from the \ClassIdx-th class, or a
one-hot vector where only the component of the predicted class is $1$
and the others are $0$.  Each training sample in the training set
\TrainDataset is an (image, label) pair, where the image might be
collected online and the label is assigned by the ML practitioner
after the data collection. The loss function in~\eqnref{eq:training}
is the cross-entropy loss~\cite{mao2023:cross}.

\subsection{Score Function} \label{sec:classifier:score_function}

Here we define the score function $\ScoreFunction{\TargetModel}$ used
in our detection algorithm for the image classifier $\TargetModel$.
The score function is a black-box membership inference test based on
the intuition that the ML model is more likely to output a confident
and correct prediction for a perturbed training sample than for a
perturbed non-training sample. This basic idea is similar to existing
label-only membership inference methods
(e.g.,~\cite{choquette2021:label}).  The confidence and correctness of
the output are measured by entropy~\cite{salem2019:ml} or modified
entropy~\cite{song2021:systematic} if the ground-truth label of
the input is known.  Specifically, we define the score function as
follows: given an input image, we first randomly generate \NumAugment
perturbed versions, and then obtain \NumAugment outputs using the
perturbed images as inputs to the target ML model.  We average the
\NumAugment outputs and use the negative (modified) entropy of the
averaged output vector elements as the score. The details of the score
function are shown in~\algref{alg:score_function_image_classifier} in
\appref{app:scorefunction}.

\subsection{Experimental Setup} \label{sec:classifier:setup}

\paragraph{Datasets} We used three image benchmarks:
CIFAR-10~\cite{krizhevsky2009:learning},
CIFAR-100~\cite{krizhevsky2009:learning}, and
TinyImageNet~\cite{Lle2015:TinyIV}:
\begin{itemize}[nosep,leftmargin=1em,labelwidth=*,align=left]
    \item \textbf{CIFAR-10}: CIFAR-10 is a dataset containing
      $60{,}000$ images of $3\times32\times32$ dimensions partitioned
      into $\NumClasses = 10$ classes. In CIFAR-10, there are
      $50{,}000$ training samples and $10{,}000$ test samples.
    \item \textbf{CIFAR-100}: CIFAR-100 is a dataset containing
      $60{,}000$ images of $3\times32\times32$ dimensions partitioned
      into $\NumClasses = 100$ classes. In CIFAR-100, there are
      $50{,}000$ training samples and $10{,}000$ test samples.
    \item \textbf{TinyImageNet}: TinyImageNet is a dataset containing
      images of $3\times64\times64$ dimensions partitioned into
      $\NumClasses = 200$ classes. In TinyImageNet, there are
      $100{,}000$ training samples and $10{,}000$ validation samples
      that we used for testing.
\end{itemize}

\paragraph{Marking setting} In each experiment, we uniformly at
random sampled \totalSamples samples
$\{\Data{\SampleIdx}\}_{\SampleIdx=1}^{\totalSamples}$ from the
training sample set \TrainingSamples of a dataset. The \totalSamples
samples are assumed to be owned by a data owner. Here we set
$\frac{\totalSamples}{\setSize{\TrainingSamples}} = 10\%$ as the
default, i.e., $\totalSamples=5{,}000$ for CIFAR-10 or CIFAR-100, and
$\totalSamples=10{,}000$ for TinyImageNet. We applied our data marking
algorithm to generate the published data
$\{\Data{\SampleIdx}{\PublishedTwin{\SampleIdx}}\}_{\SampleIdx =
  1}^{\totalSamples}$ and the unpublished data
$\{\Data{\SampleIdx}{\HiddenTwin{\SampleIdx}}\}_{\SampleIdx =
  1}^{\totalSamples}$ for
$\{\Data{\SampleIdx}\}_{\SampleIdx=1}^{\totalSamples}$. In
\eqnref{eq:twins_data}, we used $\MarkBound = 10$ as the default when
the pixel range of image is $[0, 255]$.  We defined the two marked
versions by $\Data{\SampleIdx}{0} \gets \Data{\SampleIdx} +
\Mark{\SampleIdx}$ and $\Data{\SampleIdx}{1} \gets \Data{\SampleIdx} -
\Mark{\SampleIdx}$ ($\Mark{\SampleIdx}$ is the mark), utility distance
function by $\UtilityDistance{\Data{\SampleIdx}{0}}{\Data{\SampleIdx}}
= \infinityNorm{\Data{\SampleIdx}{0} - \Data{\SampleIdx}}$ and
$\UtilityDistance{\Data{\SampleIdx}{1}}{\Data{\SampleIdx}} =
\infinityNorm{\Data{\SampleIdx}{1} - \Data{\SampleIdx}}$, and the
distance function by
$\Distance{\Data{\SampleIdx}{0}}{\Data{\SampleIdx}{1}} =
\Norm{2}{\PretrainedModel(\Data{\SampleIdx}{0}) -
  \PretrainedModel(\Data{\SampleIdx}{1})}$, where we used
ResNet18~\cite{he2016:deep} pretrained on
ImageNet~\cite{deng2009:imagenet} to be the default feature extractor
$\PretrainedModel$.  We solved \eqnref{eq:twins_data} by projected
gradient descent~\cite{lin2007:projected}.  Then we uniformly at
random sampled a subset (of size $\PublishedInTrainDatasetSize$) of
$\{\Data{\SampleIdx}{\PublishedTwin{\SampleIdx}}\}_{\SampleIdx =
  1}^{\totalSamples}$ as $\PublishedInTrainDataset$ (i.e.,
$\PublishedInTrainDataset \subseteq
\{\Data{\SampleIdx}{\PublishedTwin{\SampleIdx}}\}_{\SampleIdx =
  1}^{\totalSamples}$) to simulate a general case where the ML
practitioner collected a subset of published data as training samples.
By default, we set $\PublishedInTrainDatasetSize = \totalSamples$. As
such, we constituted the training dataset collected by the ML
practitioner as $\TrainDataset =(\TrainingSamples \setminus
\{\Data{\SampleIdx}\}_{\SampleIdx=1}^{\totalSamples}) \cup
\PublishedInTrainDataset$ with correct labels (i.e., using the same
labels as those in the dataset). Some examples of marked images are
displayed in \figref{fig:cifar10_examples},
\figref{fig:cifar100_examples}, and \figref{fig:tinyimagenet_examples}
in \appref{app:marked_examples}.

\paragraph{Training setting} We used ResNet18 as the default
architecture of the ML model \TargetModel trained by the ML
practitioner.  We used a standard SGD algorithm to train \TargetModel,
as follows: \TargetModel was trained on normalized training data with
default data augmentation applied~\cite{geiping2020:witches} using an
SGD optimizer~\cite{amari1993:backpropagation} with a weight decay of
$5\times10^{-4}$ for 80 epochs, a batch size of 128, and an initial
learning rate of 0.1 decayed by a factor of 0.1 when the number of
epochs reached 30, 50, or 70.

\paragraph{Detection setting} In each detection experiment, we
applied our data-use detection algorithm to the given ML model
\TargetModel using a set of pairs of generated published data and
unpublished data. In the data-use detection algorithm and the score
function, we set $\Confidence=0.025$, $\FalseDetectionRate=0.05$,
and $\NumAugment=16$ as the default.  (Recall from
\thmref{theorem:fdr} that \FalseDetectionRate bounds the
false-detection rate.)  We present results for four different
experimental conditions that define the information available to the
detector, denoted as $\Knowledge{1}$, $\Knowledge{2}$,
$\Knowledge{3}$, and $\Knowledge{4}$.  We define these four conditions
in \tblref{table:scenarios}.
\begin{table}[t]
    \centering
    \begin{tabular}{ccc}
    \toprule
    \textbf{Condition} & \textbf{Confidence score} & \textbf{Ground-truth} \\
    \midrule
    $\Knowledge{1}$ & \chkmrk & \chkmrk\\
    $\Knowledge{2}$ & \chkmrk & \xmark \\
    $\Knowledge{3}$ & \xmark & \chkmrk \\
    $\Knowledge{4}$ & \xmark & \xmark \\
    \bottomrule
    \end{tabular}
    \vspace{1ex}
    \caption{Information available to the detector.
      ``\textbf{Confidence score}'' indicates whether the ML model
      \TargetModel outputs a full confidence vector (``\chkmrk'') or
      just a label, i.e., a one-hot vector (``\xmark'').
      ``\textbf{Ground-truth}'' indicates whether the true label of a
      query to the ML model is known by the detector (``\chkmrk'') or
      not (``\xmark'').}
    \label{table:scenarios}
    \vspace{-4ex}
\end{table}

\paragraph{Baselines} We used two state-of-the-art methods,
Radioactive Data~\cite{sablayrolles2020:radioactive}, which we
abbreviate to RData, and Untargeted Backdoor Watermark-Clean
(UBW-C)~\cite{li2022:untargeted}, as baselines. RData requires
knowledge of the class labels for its data.  So, we also consider two
variants of RData in which the data owner is presumed to not know how
the ML practitioner will label her data, and so applies the same mark
to all of her data regardless of class (``RData (one mark)''), or to
know only a ``coarse'' label (superclass) of the class label the ML
practitioner will assign to each (``RData (superclass)'').  The
details of baselines and their implementation are described in
\appref{app:baselines}.

\paragraph{Metrics}
We used the following metrics to evaluate the methods:
\begin{itemize}[nosep,leftmargin=1em,labelwidth=*,align=left]
  \item \textbf{Test accuracy} (\Accuracy): \Accuracy is the fraction
    of test samples that are correctly classified by the ML model
    \TargetModel. A higher \Accuracy indicates a better performance of
    the ML model.
 \item \textbf{Detection success rate} ($\DetectionSuccessRate$):
   $\DetectionSuccessRate$ is the fraction of detection experiments
   returning True (i.e., affirmatively detecting data use). When the
   detected ML model did use the published data, a higher
   $\DetectionSuccessRate$ indicates a better performance of the data
   auditing framework.  When the detected ML model did not use the
   published data, a lower $\DetectionSuccessRate$ indicates more
   robustness to false detections.
 \item \textbf{Minimum amount of published data used in training, as a
   percentage of the training data set, to trigger
   detection} ($\MinimalPercentage$): That is, \MinimalPercentage is
   the minimum value of $\PublishedInTrainDatasetSize /
   \TrainDatasetSize$, expressed as a percentage, at which
   the detection algorithm returns True.  Therefore, a lower
   $\MinimalPercentage$ indicates a more sensitive detector.  However,
   to find $\MinimalPercentage$ in each of our settings is costly
   since we need to exhaustively test potential values of
   $\PublishedInTrainDatasetSize / \TrainDatasetSize$.
   For this reason, we report an alternative measure (see below) in
   place of $\MinimalPercentage$.
 \item \textbf{Query cost} ($\NumQuery$): $\NumQuery$ is the number of
   queries to the target ML model \TargetModel to conclude that
   \TargetModel was trained on the data-owner's data.  That is,
   $\NumQuery = 2 \times \NumAugment \times \NumberQueriedPublished$,
   where $\NumberQueriedPublished$ ($\NumberQueriedPublished\leq
   \totalSamples$) is the number of published data used to query the
   ML model to detect its training with the data-owner's data. It
   indicates the practical cost used in the detection step. A lower
   \NumQuery indicates a more cost-efficient detection method.
 \item \textbf{Ratio between the number of queried published data and
   the total number of training samples}
   ($\frac{\NumberQueriedPublished}{\TrainDatasetSize}$):
   $\frac{\NumberQueriedPublished}{\TrainDatasetSize}$ is the
   ratio between the number $\NumberQueriedPublished$ of published
   data used to query the ML model (resulting in detection) and the
   total number $\TrainDatasetSize$ of training samples.  In our
   tests, $\frac{\NumberQueriedPublished}{\TrainDatasetSize}$
   was strongly correlated with $\MinimalPercentage$ (Pearson
   correlation coefficient~\cite{cohen2009:pearson} $=~0.66$ with high statistical
   significance; see~\appref{app:correlation}) and is considerably
   cheaper to compute than $\MinimalPercentage$. Moreover, for a fixed
   $\NumAugment$ and $\TrainDataset$, $\NumQuery$ is a linear function
   of $\frac{\NumberQueriedPublished}{\TrainDatasetSize}$.
   Therefore, when presenting our results, we use
   $\frac{\NumberQueriedPublished}{\TrainDatasetSize}$ as a
   surrogate for $\MinimalPercentage$ and $\NumQuery$.  A lower
   $\frac{\NumberQueriedPublished}{\TrainDatasetSize}$ indicates
   a lower $\MinimalPercentage$ and a lower $\NumQuery$, and thus it
   suggests a more detection-efficient and more cost-efficient method.
\end{itemize}

\subsection{Experimental Results} \label{sec:classifier:results}

\subsubsection{Overall Performance} \label{sec:classifier:results:overall}

\paragraph{Effectiveness} The detection performance of our proposed
method on different visual benchmarks is shown in
\tblref{table:classifier:overview}.
\tblref{table:classifier:overview} demonstrates that our method is
highly effective to detect the use of published data in training ML
models, i.e, yielding a $20/20$ $\DetectionSuccessRate$ in all
settings where the published data is used as a subset of training
samples of the target ML model. In addition, the ML models trained on
the datasets including the published data preserved good utility,
i.e., their $\Accuracy$ values are only slightly lower ($<1\%$ on
average) than those trained on clean datasets. For detection, we
needed a $\frac{\NumberQueriedPublished}{\TrainDatasetSize}$
ranging from $2.20\%$ to $4.65\%$ for CIFAR-10, from $0.19\%$ to
$0.60\%$ for CIFAR-100, and from $0.14\%$ to $0.67\%$ for
TinyImageNet. These results show that our method achieved more
detection efficiency when applied to a classification task with a
large number of classes. Such ranges of
$\frac{\NumberQueriedPublished}{\TrainDatasetSize}$ also
indicate that detection needs a number of queries to the ML model (i.e.,
$\NumQuery$) ranging from a hundred to tens of thousands. Given the
current prices of online queries to pretrained visual AI models (e.g.,
$\$1.50$ per $1{,}000$
images\footnote{https://cloud.google.com/vision/pricing}), the
detection cost is affordable, ranging from several dollars to a
hundred dollars. When we have less information on the output of the ML
model (i.e., the outputs are the predictions only) or the queries
(i.e., the ground-truth labels are unknown) in the detection, we
needed more queries to trigger detection, i.e., yielding a
larger $\frac{\NumberQueriedPublished}{\TrainDatasetSize}$.

\begin{table*}[ht!]
    \centering
    \begin{tabular}{@{}lr@{\hspace{0.5em}}rr@{\hspace{0.5em}}rr@{\hspace{0.5em}}rr@{\hspace{0.5em}}rr@{\hspace{0.5em}}rr@{\hspace{0.5em}}rr@{\hspace{0.5em}}rr@{\hspace{0.5em}}r@{}}
    \toprule
    \multicolumn{1}{c}{} & \multicolumn{1}{c}{\multirow{2}{*}{$\Accuracy \%$}}& \multicolumn{1}{c}{\multirow{2}{*}{$\AccuracyDifference \%$}}& \multicolumn{2}{c}{$\Knowledge{1}$}& \multicolumn{2}{c}{$\Knowledge{2}$}& \multicolumn{2}{c}{$\Knowledge{3}$}& \multicolumn{2}{c}{$\Knowledge{4}$}\\ 
    & & & \multicolumn{1}{c}{$\DetectionSuccessRate$} & \multicolumn{1}{c}{$\frac{\NumberQueriedPublished}{\TrainDatasetSize}$} & \multicolumn{1}{c}{$\DetectionSuccessRate$} & \multicolumn{1}{c}{$\frac{\NumberQueriedPublished}{\TrainDatasetSize}$} & \multicolumn{1}{c}{$\DetectionSuccessRate$} & \multicolumn{1}{c}{$\frac{\NumberQueriedPublished}{\TrainDatasetSize}$} & \multicolumn{1}{c}{$\DetectionSuccessRate$} & \multicolumn{1}{c}{$\frac{\NumberQueriedPublished}{\TrainDatasetSize}$} \\ 
    \midrule 
    CIFAR-10 & $93.64$ & $-0.05$ & $20/20$ & $2.20\%$ &  $20/20$ & $2.67\%$ & $20/20$ & $4.22\%$ & $20/20$ & $4.65\%$ \\
    CIFAR-100 & $74.29$ & $-0.76$ & $20/20$ & $0.19\%$ &  $20/20$ & $0.20\%$ & $20/20$ & $0.59\%$ & $20/20$ & $0.60\%$ \\
    TinyImageNet & $59.13$ & $-0.16$ & $20/20$ & $0.14\%$ &  $20/20$ & $0.13\%$ & $20/20$ & $0.59\%$ & $20/20$ & $0.67\%$ \\ 

    \bottomrule 
    \end{tabular}
    \vspace{1ex}
    \caption{Overall performance of our proposed method on different
      image benchmarks, with an upper bound of $\FalseDetectionRate =
      0.05$ on the false-detection rate.  All results are averaged
      over 20 experiments.  The numbers in the $\AccuracyDifference\%$
      column are the differences between averaged accuracies of ML
      models trained on marked datasets and those of ML models trained
      on clean datasets.}
    \vspace{-4ex} 
    \label{table:classifier:overview}    
\end{table*}

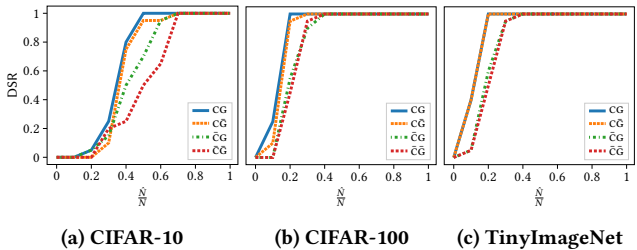
\begin{figure}
    \centering
    \begin{subfigure}[b]{0.176\textwidth}
        \setlength\figureheight{1.9in}
        \centering
        \resizebox{\textwidth}{!}{
\begin{tikzpicture}[font=\huge]

\definecolor{crimson2143940}{RGB}{214,39,40}
\definecolor{darkgray176}{RGB}{176,176,176}
\definecolor{darkorange25512714}{RGB}{255,127,14}
\definecolor{forestgreen4416044}{RGB}{44,160,44}
\definecolor{lightgray204}{RGB}{204,204,204}
\definecolor{steelblue31119180}{RGB}{31,119,180}

\begin{axis}[
  legend cell align={left},
  legend style={
    fill opacity=0.8,
    draw opacity=1,
    text opacity=1,
    at={(0.97,0.03)},
    anchor=south east,
    draw=lightgray204
  },
  tick align=outside,
  tick pos=left,
  x grid style={darkgray176},
  xlabel={$\frac{\PublishedInTrainDatasetSize}{\totalSamples}$},
  xmin=-0.055, xmax=1.045,
  xtick style={color=black},
  y grid style={darkgray176},
  ylabel={$\DetectionSuccessRate$},
  ymin=-0.05, ymax=1.05,
  ytick style={color=black}
  ]
\addplot [line width=3, steelblue31119180]
table {%
0 0
0.1 0
0.2 0.05
0.3 0.25
0.4 0.8
0.5 1
0.6 1
0.7 1
0.8 1
0.9 1
1 1
};
\addlegendentry{$\Knowledge{1}$}
\addplot [line width=3, densely dotted, darkorange25512714]
table {%
0 0
0.1 0
0.2 0
0.3 0.1
0.4 0.75
0.5 0.95
0.6 0.95
0.7 1
0.8 1
0.9 1
1 1
};
\addlegendentry{$\Knowledge{2}$}
\addplot [line width=3, dash pattern=on 1pt off 3pt on 3pt off 3pt, forestgreen4416044]
table {%
0 0
0.1 0
0.2 0.05
0.3 0.15
0.4 0.5
0.5 0.7
0.6 0.95
0.7 1
0.8 1
0.9 1
1 1
};
\addlegendentry{$\Knowledge{3}$}
\addplot [line width=3, dotted, crimson2143940]
table {%
0 0
0.1 0
0.2 0
0.3 0.2
0.4 0.25
0.5 0.5
0.6 0.65
0.7 1
0.8 1
0.9 1
1 1
};
\addlegendentry{$\Knowledge{4}$}
\end{axis}

\end{tikzpicture}}
        \caption{CIFAR-10}
      \end{subfigure}
      \hfill
      \begin{subfigure}[b]{0.144\textwidth}
        \setlength\figureheight{1.9in}
        \centering
        \resizebox{\textwidth}{!}{
\begin{tikzpicture}[font=\huge]

\definecolor{crimson2143940}{RGB}{214,39,40}
\definecolor{darkgray176}{RGB}{176,176,176}
\definecolor{darkorange25512714}{RGB}{255,127,14}
\definecolor{forestgreen4416044}{RGB}{44,160,44}
\definecolor{lightgray204}{RGB}{204,204,204}
\definecolor{steelblue31119180}{RGB}{31,119,180}

\begin{axis}[
    legend cell align={left},
    legend style={
      fill opacity=0.8,
      draw opacity=1,
      text opacity=1,
      at={(0.97,0.03)},
      anchor=south east,
      draw=lightgray204
    },
    tick align=outside,
    tick pos=left,
    x grid style={darkgray176},
    xlabel={$\frac{\PublishedInTrainDatasetSize}{\totalSamples}$},
    xmin=-0.055, xmax=1.045,
    xtick style={color=black},
    y grid style={darkgray176},
    ymajorticks=false,
    ymin=-0.05, ymax=1.05,
    ytick style={color=black}
    ]
\addplot [line width=3, steelblue31119180]
table {%
0 0
0.1 0.25
0.2 1
0.3 1
0.4 1
0.5 1
0.6 1
0.7 1
0.8 1
0.9 1
1 1
};
\addlegendentry{$\Knowledge{1}$}
\addplot [line width=3, densely dotted, darkorange25512714]
table {%
0 0
0.1 0.1
0.2 0.95
0.3 1
0.4 1
0.5 1
0.6 1
0.7 1
0.8 1
0.9 1
1 1
};
\addlegendentry{$\Knowledge{2}$}
\addplot [line width=3,  dash pattern=on 1pt off 3pt on 3pt off 3pt,  forestgreen4416044]
table {%
0 0
0.1 0
0.2 0.55
0.3 0.9
0.4 1
0.5 1
0.6 1
0.7 1
0.8 1
0.9 1
1 1
};
\addlegendentry{$\Knowledge{3}$}
\addplot [line width=3, dotted, crimson2143940]
table {%
0 0
0.1 0
0.2 0.45
0.3 0.95
0.4 1
0.5 1
0.6 1
0.7 1
0.8 1
0.9 1
1 1
};
\addlegendentry{$\Knowledge{4}$}
\end{axis}

\end{tikzpicture}}
        \caption{CIFAR-100}
      \end{subfigure}
      \hfill
      \begin{subfigure}[b]{0.144\textwidth}
        \setlength\figureheight{1.9in}
        \centering
        \resizebox{\textwidth}{!}{
\begin{tikzpicture}[font=\huge]

    \definecolor{crimson2143940}{RGB}{214,39,40}
    \definecolor{darkgray176}{RGB}{176,176,176}
    \definecolor{darkorange25512714}{RGB}{255,127,14}
    \definecolor{forestgreen4416044}{RGB}{44,160,44}
    \definecolor{lightgray204}{RGB}{204,204,204}
    \definecolor{steelblue31119180}{RGB}{31,119,180}
    
    \begin{axis}[
        legend cell align={left},
        legend style={
          fill opacity=0.8,
          draw opacity=1,
          text opacity=1,
          at={(0.97,0.03)},
          anchor=south east,
          draw=lightgray204
        },
        tick align=outside,
        tick pos=left,
        x grid style={darkgray176},
        xlabel={$\frac{\PublishedInTrainDatasetSize}{\totalSamples}$},
        xmin=-0.055, xmax=1.045,
        xtick style={color=black},
        y grid style={darkgray176},
        ymajorticks=false,
        ymin=-0.05, ymax=1.05,
        ytick style={color=black}
        ]
\addplot [line width=3, steelblue31119180]
table {%
0 0
0.1 0.4
0.2 1
0.3 1
0.4 1
0.5 1
0.6 1
0.7 1
0.8 1
0.9 1
1 1
};
\addlegendentry{$\Knowledge{1}$}
\addplot [line width=3, densely dotted, darkorange25512714]
table {%
0 0
0.1 0.4
0.2 1
0.3 1
0.4 1
0.5 1
0.6 1
0.7 1
0.8 1
0.9 1
1 1
};
\addlegendentry{$\Knowledge{2}$}
\addplot [line width=3,  dash pattern=on 1pt off 3pt on 3pt off 3pt,  forestgreen4416044]
table {%
0 0
0.1 0.05
0.2 0.6
0.3 0.95
0.4 1
0.5 1
0.6 1
0.7 1
0.8 1
0.9 1
1 1
};
\addlegendentry{$\Knowledge{3}$}
\addplot [line width=3, dotted, crimson2143940]
table {%
0 0
0.1 0.05
0.2 0.5
0.3 0.95
0.4 1
0.5 1
0.6 1
0.7 1
0.8 1
0.9 1
1 1
};
\addlegendentry{$\Knowledge{4}$}
\end{axis}

\end{tikzpicture}}
        \caption{TinyImageNet}
      \end{subfigure}
      \vspace{-2ex}
    \caption{The impact of $\frac{\PublishedInTrainDatasetSize}{\totalSamples}$ on the detection performance 
            (the default $\frac{\PublishedInTrainDatasetSize}{\totalSamples}$ is $1.0$). The results from 
            $\frac{\PublishedInTrainDatasetSize}{\totalSamples} = 0$ are the false-detections of our method.}
    \vspace{-2ex}
    \label{fig:used_published}
\end{figure}

\paragraph{Impact of using published data partially and
  false-detections} After the published data is released online, the
ML practitioner might collect them partially (i.e.,
$\frac{\PublishedInTrainDatasetSize}{\totalSamples}$ is smaller than
$1.0$) and use the collected data in training.  Here, we tested the
detection performance of our method on the ML model trained on
$\TrainDataset$ under different ratios of
$\frac{\PublishedInTrainDatasetSize}{\totalSamples}$.  The results are
shown in \figref{fig:used_published}. When the ML practitioner used
more published data, $\DetectionSuccessRate$ was higher. Especially,
when he used $\geq 70\%$ published CIFAR-10 data, or $\geq 40\%$
published CIFAR-100 data or published TinyImageNet data, we achieved a
$\DetectionSuccessRate$ of $20/20$, even with the least information
(condition $\Knowledge{4}$).  When the ML practitioner did not use any
published data in training (i.e.,
$\frac{\PublishedInTrainDatasetSize}{\totalSamples} = 0$),
$\DetectionSuccessRate$ was $0/20$ under all considered settings,
which empirically confirms the upper bound $\FalseDetectionRate =
0.05$ on false-detection rate of our method.

\paragraph{Comparison with baselines}
\tblref{table:classifier:baselines} summarizes the comparison between
our method and baselines.  Compared with the baselines, our method is
more effective in the detection of data use, i.e., yielding a higher
\DetectionSuccessRate and a higher \Accuracy. More importantly,
different from the two state-of-the-art methods (i.e., RData and
UBW-C), our method does not need the labeling of training samples
before data publication or the white-box access to the ML model (i.e.,
knowing the parameters of the ML model).  The variants of RData
denoted as ``one mark'' and ``superclass'' do not need the complete
information on labeling, but their $\DetectionSuccessRate$ dropped
significantly.

\begin{table*}[ht!]
    \centering
    \begin{tabular}{@{}lr@{\hspace{0.8em}}cc@{\hspace{0.8em}}cc@{\hspace{0.8em}}cc@{\hspace{0.8em}}cc@{\hspace{0.8em}}cc@{\hspace{0.8em}}cc@{\hspace{0.8em}}cc@{\hspace{0.8em}}r@{}}
    \toprule
    \multicolumn{1}{c}{} & \multicolumn{1}{c}{} & Labeling & White & Bounded & \multicolumn{2}{c}{$1\%$} & \multicolumn{2}{c}{$2\%$}& \multicolumn{2}{c}{$5\%$}& \multicolumn{2}{c}{$10\%$}\\ 
    & & known & box & FDR & \multicolumn{1}{c}{$\DetectionSuccessRate$}& \multicolumn{1}{c}{\Accuracy $\%$} & \multicolumn{1}{c}{$\DetectionSuccessRate$}& \multicolumn{1}{c}{\Accuracy $\%$}& \multicolumn{1}{c}{$\DetectionSuccessRate$}& \multicolumn{1}{c}{\Accuracy $\%$}& \multicolumn{1}{c}{$\DetectionSuccessRate$}& \multicolumn{1}{c}{\Accuracy $\%$}\\ 
    \midrule 
    \multirow{5}{*}{CIFAR-10} & Our method ($\Knowledge{4}$) & \emptycircle{} & \emptycircle{} & \chkmrk & $\boldsymbol{8/20}$ & $\boldsymbol{93.79}$ & $\boldsymbol{11/20}$ & $\boldsymbol{93.71}$ & $\boldsymbol{19/20}$ & $\boldsymbol{93.70}$ & $\boldsymbol{20/20}$ & $\boldsymbol{93.64}$ \\ 
    & RData& \fullcircle{} & \fullcircle{} & \chkmrk & $1/20$ & $93.65$ & $2/20$ & $93.56$ & $2/20$ & $93.29$ & $4/20$ & $93.26$ \\ 
    & RData (one mark)& \emptycircle{} & \fullcircle{} & \chkmrk & $0/20$ & $93.75$ & $0/20$ & $93.60$ & $0/20$ & $93.42$ & $0/20$ & $93.25$ \\ 
    & UBW-C ($\UBWCParameter=0.25$) & \fullcircle{} & \emptycircle{} & \xmark & $0/20$ & $93.50$ & $0/20$ & $93.14$ & $0/20$ & $92.67$ & $2/20$ & $92.73$ \\
    & UBW-C ($\UBWCParameter=0.20$) & \fullcircle{} & \emptycircle{} & \xmark & $1/20$ & $93.50$ & $8/20$ & $93.15$ & $7/20$ & $92.46$ & $15/20$ & $92.52$ \\ 
    \midrule 
    \multirow{6}{*}{CIFAR-100} & Our method ($\Knowledge{4}$) & \emptycircle{} & \emptycircle{} & \chkmrk & $\boldsymbol{20/20}$ & $\boldsymbol{75.01}$ & $\boldsymbol{20/20}$ & $\boldsymbol{74.94}$ & $\boldsymbol{20/20}$ & $\boldsymbol{74.60}$ & $\boldsymbol{20/20}$ & $\boldsymbol{74.29}$\\ 
    & RData &\fullcircle{} & \fullcircle{} & \chkmrk & $5/20$ & $74.66$ & $14/20$ & $74.57$ & $\boldsymbol{20/20}$ & $73.81$ & $\boldsymbol{20/20}$ & $73.53$ \\ 
    & RData (superclass)& \halfcircle{} & \fullcircle{} & \chkmrk & $4/20$ & $74.76$ & $10/20$ & $74.46$ & $14/20$ & $73.99$ & $19/20$ & $73.42$ \\ 
    & RData (one mark)& \emptycircle{} & \fullcircle{} & \chkmrk & $0/20$ & $74.70$ & $0/20$ & $74.51$ & $1/20$ & $74.05$ & $0/20$ & $73.51$ \\ 
    & UBW-C ($\UBWCParameter=0.25$) & \fullcircle{} & \emptycircle{} & \xmark & $0/20$ & $74.60$ & $0/20$ & $74.16$ & $16/20$ & $73.30$ & $\boldsymbol{20/20}$ & $72.32$ \\\
    & UBW-C ($\UBWCParameter=0.20$)& \fullcircle{} & \emptycircle{} & \xmark & $19/20$ & $74.60$ & $\boldsymbol{20/20}$ & $74.33$ & $\boldsymbol{20/20}$ & $73.21$ & $\boldsymbol{20/20}$ & $72.47$ \\
    \midrule 
    \multirow{6}{*}{TinyImageNet} & Our method ($\Knowledge{4}$) &\emptycircle{} &\emptycircle{} & \chkmrk & $\boldsymbol{20/20}$ & $\boldsymbol{59.32}$ & $\boldsymbol{20/20}$ & $\boldsymbol{59.24}$ & $\boldsymbol{20/20}$ & $\boldsymbol{59.17}$ & $\boldsymbol{20/20}$ & $\boldsymbol{59.13}$ \\ 
    & RData&\fullcircle{} & \fullcircle{}& \chkmrk & $8/20$ & $59.14$ & $18/20$ & $58.94$ & $\boldsymbol{20/20}$ & $58.59$ & $\boldsymbol{20/20}$ & $58.13$ \\ 
    & RData (superclass)& \halfcircle{} & \fullcircle{}& \chkmrk & $7/20$ & $59.14$ & $14/20$ & $59.03$ & $\boldsymbol{20/20}$ & $58.71$ & $\boldsymbol{20/20}$ & $58.09$ \\
    & RData (one mark)&\emptycircle{} & \fullcircle{}& \chkmrk & $2/20$ & $59.12$ & $1/20$ & $58.98$ & $0/20$ & $58.61$ & $0/20$ & $58.29$ \\ 
    & UBW-C ($\UBWCParameter=0.25$) & \fullcircle{} & \emptycircle{} & \xmark & $0/20$ & $59.01$ & $0/20$ & $58.80$ & $0/20$ & $58.43$ & $0/20$ & $57.78$ \\
    & UBW-C ($\UBWCParameter=0.20$)& \fullcircle{} & \emptycircle{} & \xmark & $0/20$ & $59.01$ & $0/20$ & $58.62$ & $6/20$ & $58.41$ & $17/20$ & $57.63$ \\ 
    \bottomrule 
    \end{tabular}
    \vspace{1ex}
    \caption[Short caption without TikZ commands]{Comparison between
      our proposed method and baselines under different rates of
      $\frac{\PublishedInTrainDatasetSize}{\TrainDatasetSize} \in
      \{1\%, 2\%, 5\%, 10\%\}$.  The results of our method come from
      the setting with least information available to the data owner,
      i.e., $\Knowledge{4}$. In UBW-C, $\UBWCParameter$ is a
      hyperparameter of its detection algorithm. In the columns of
      ``Labeling known'' and ``White box'', ``\fullcircle{}''
      indicates that the information is needed; ``\emptycircle{}''
      means that information is not needed; ``\halfcircle{}'' means
      that partial information is needed. In the column ``bounded
      FDR'', ``\chkmrk'' (``\xmark'') indicates that the method
      provides (does not provide) a provable bound on the
      false-detection rate.  Results are averaged over 20
      experiments. The bold results are the best ones among the
      compared methods.}
    \vspace{-4ex}
    \label{table:classifier:baselines}    
\end{table*}

\paragraph{Multiple data owners} Here we consider a general
real-world setting where there are multiple data owners applying data
auditing independently, each of which set the upper bound on the
false-detection rate as $\FalseDetectionRate = 0.05$.  In these
experiments, each data owner had $5{,}000$ CIFAR-100 data items (i.e.,
$10\%$ of the training samples collected by the ML practitioner) to
publish. Each applied an auditing framework to generate her marked
data and to detect its use in the deployed ML model independently.
The detection results with our method and with the state-of-the-art
method, RData (with full information on data labeling), are shown in
\tblref{table:classifier:multiple_users}.  Compared with RData, whose
detection performance degraded with a larger number of data owners,
our method was much more effective, yielding a $100\%$
$\DetectionSuccessRate$ in all cases.

\begin{table}[ht!]
  \vspace{2ex}
    \centering
    \begin{tabular}{@{}lrrrr@{}}
    \toprule
    & \multicolumn{4}{c}{Data owners} \\
    & \multicolumn{1}{c}{1} & \multicolumn{1}{c}{2} & \multicolumn{1}{c}{5} & \multicolumn{1}{c}{10}\\  
    \midrule 
    Our method ($\Knowledge{4}$) & $20/20$ & $40/40$ & $100/100$ & $200/200$  \\ 
    RData & $20/20$ & $38/40$ & $64/100$ & $90/200$  \\ 
    \bottomrule 
    \end{tabular}
    \vspace{1ex}
    \caption{Comparison between our method and RData (which requires
      knowledge of data labeling), both under an upper bound of
      $\FalseDetectionRate = 0.05$ on the false-detection rate, when
      multiple data owners applied data auditing independently.  Each
      owner contributed $10\%$ of the training dataset.  Results are
      the total detections over all detection attempts (by all data
      owners) in $20$ experiments.}
    \vspace{-4ex}
    \label{table:classifier:multiple_users}    
\end{table}

The results in \tblref{table:classifier:overview},
\figref{fig:used_published}, \tblref{table:classifier:baselines},
\tblref{table:classifier:multiple_users} demonstrate that our method
achieves our \textit{effectiveness} goal defined in
\secref{sec:problem:design_goal}.  \tblref{table:classifier:baselines}
and \tblref{table:classifier:multiple_users} show the advantages of
our proposed method over the baselines.
\tblref{table:classifier:multiple_users} presents interesting results
under real-world settings where multiple data owners independently
audit an ML model for use of their data.

\subsubsection{Impact of ML Model Architecture and Hyperparameters}
\label{sec:classifier:results:hyperparameters}
In this section, we explore the impact of the ML practitioner's model
architecture and the data owner's hyperparameters on detection, such
as the utility bound $\MarkBound$, the feature extractor
$\PretrainedModel$ used to generate marked data, the upper bound
\FalseDetectionRate on the false-detection rate, and the number
$\NumAugment$ of sampled perturbations per image in detection. Due to
the space limit, we present results in \appref{app:hyperparameters}.

\subsubsection{Robustness to Countermeasures/Adaptive Attacks}
\label{sec:classifier:results:adaptive_attack}

When the ML practitioner knows that a data owner marked her data, he
might utilize countermeasures/adaptive attacks to defeat the auditing
method. His goal is to decrease $\DetectionSuccessRate$ without
degrading the performance of the trained ML model significantly.  We
evaluated the robustness of the proposed method to three types of
countermeasures/adaptive attacks, described below.

\paragraph{Limiting the information from the ML model output} Since
our detection method measures the difference between outputs of the ML
model on the published data and unpublished data, the ML practitioner
can limit the output (e.g., the vector of confidence scores) of the
deployed ML model, aiming to degrade our detection. Here we considered
two countermeasures of this type:
\begin{itemize}[nosep,leftmargin=1em,labelwidth=*,align=left]
    \item Outputting only the top $\TopK$ confidence scores
      ($\TopScores{\TopK}$): This countermeasure allows the deployed
      ML model to output the top $\TopK$ confidence scores, masking
      out the others in the output vector. Here we considered
      $\TopK=1$ and $\TopK=5$.
    \item Adding perturbation into the output
      ($\MemGuard$~\cite{jia2019:memguard}): This countermeasure adds
      carefully crafted perturbations into the ML model output to
      limit the information given. We considered $\MemGuard$ proposed
      by Jia, et al.~\cite{jia2019:memguard} to design the
      perturbation, where we used a moderate distortion budget of
      $0.5$.
\end{itemize}
Note that these countermeasures can be applied only in the ML model
deployment where the output is a vector of confidence scores instead
of a prediction/label.

\paragraph{Reducing memorization of training samples} Intuitively,
the ML practitioner can apply methods to discourage memorization of
training samples by the ML model, so that the published data and the
unpublished data will have similar scores by the defined score
function. As
such, reducing memorization of training samples could render the
detection method to be less effective.  We considered three such
countermeasures:
\begin{itemize}[nosep,leftmargin=1em,labelwidth=*,align=left]
    \item Differential privacy
      ($\DifferentialPrivacy$~\cite{dwork2006:differential}):
      $\DifferentialPrivacy$ is a standard privacy definition that
      limits the information leaked about any training input in the
      output of the algorithm. To achieve $\DifferentialPrivacy$, the
      ML practitioner clips the gradients of each training batch and
      adds Gaussian noise (with standard deviation of
      $\GaussianNoiseParameter$) into the clipped gradients during ML
      model training~\cite{abadi2016:deep}.
    \item Early stopping ($\EarlyStopping$): In this countermeasure,
      the ML practitioner trains the ML model for a small number of
      epochs to prevent the ML model from overfitting to the training
      samples. Here we trained ML models for $20$, $40$, and $60$
      epochs, denoted as $\EarlyStopping(20)$, $\EarlyStopping(40)$,
      and $\EarlyStopping(60)$, respectively.
    \item Adversarial regularization
      ($\AdversarialRegularization$~\cite{nasr2018:machine}):
      Adversarial regularization is a strategy to generalize the ML
      model.  It does so by alternating between training the ML model
      to minimize the classification loss and training it to maximize
      the gain of a membership inference attack. In the implementation
      of $\AdversarialRegularization$, we set the adversarial
      regularization factor to be $1.0$~\cite{nasr2018:machine}.
\end{itemize}

\paragraph{Other attacks} We also considered some other adaptive
attacks that aim to defeat our auditing method:
\begin{itemize}[nosep,leftmargin=1em,labelwidth=*,align=left]
    \item Detecting pairs of published data and unpublished data in
      queries ($\TwinsDetection$): The intuition behind this pair
      detection is that if the deployment can detect queries of a pair
      of published data and unpublished data, then it will return the
      same output to evade detection.  We design such a pair detection
      method as follows: we maintain a window of queries in the
      history and their ML model outputs, and we compare each new
      query with those in the window to decide what to output.  If the
      infinity norm of the pixel difference between the new query and
      a previous query is smaller than $2\MarkBound$, we return the
      output of the previous query; otherwise, we return the output
      for the new query.
    \item Adding Gaussian noise into the training samples
      ($\GaussianNoise(\GaussianNoiseParameter)$): This method adds
      noise into each training sample to mask the added mark. The
      added noise is sampled from a Gaussian distribution
      with standard deviation $\GaussianNoiseParameter$.
    \item Avoiding data augmentation in training
      ($\NoTrainAugmentation$): Excluding data augmentation in ML
      model training will degrade the effectiveness of the label-only
      membership inference that we apply as the score function for the
      image classifier, as demonstrated by previous works
      (e.g.,~\cite{choquette2021:label}).
    \item Using our marking algorithm (with the default
      hyperparameters) to perturb training samples ($\MarkPerturb$):
      This countermeasure applies our marking algorithm (with the
      default hyperparameters) to craft two perturbed versions of each
      training sample and randomly selects one to use in
      training.\footnote{We could use both perturbed versions in
      training but we would need to reduce the number of epochs to
      half (i.e., 40 epochs) for fair comparison.}
\end{itemize}

\begin{table*}[ht!]
    \centering
    \begin{tabular}{@{}lr@{\hspace{0.5em}}rr@{\hspace{0.5em}}rr@{\hspace{0.5em}}rr@{\hspace{0.5em}}rr@{\hspace{0.5em}}rr@{\hspace{0.5em}}rr@{\hspace{0.5em}}r@{}}
    \toprule
    \multicolumn{1}{c}{} & \multicolumn{1}{c}{} & \multicolumn{1}{c}{\multirow{2}{*}{$\Accuracy \%$}}& \multicolumn{2}{c}{$\Knowledge{1}$}& \multicolumn{2}{c}{$\Knowledge{2}$}& \multicolumn{2}{c}{$\Knowledge{3}$}& \multicolumn{2}{c}{$\Knowledge{4}$}\\ 
    & & & \multicolumn{1}{c}{$\DetectionSuccessRate$} & \multicolumn{1}{c}{$\frac{\NumberQueriedPublished}{\TrainDatasetSize}$} & \multicolumn{1}{c}{$\DetectionSuccessRate$} & \multicolumn{1}{c}{$\frac{\NumberQueriedPublished}{\TrainDatasetSize}$} & \multicolumn{1}{c}{$\DetectionSuccessRate$} & \multicolumn{1}{c}{$\frac{\NumberQueriedPublished}{\TrainDatasetSize}$} & \multicolumn{1}{c}{$\DetectionSuccessRate$} & \multicolumn{1}{c}{$\frac{\NumberQueriedPublished}{\TrainDatasetSize}$} \\ 
    
    \midrule 
    No adaptive attack & & $74.29$ & $20/20$ & $0.19\%$ & $20/20$ & $0.20\%$ &  $20/20$ & $0.59\%$ & $20/20$ & $0.60\%$ \\
    \midrule 
    \multirow{3}{*}{Masking output} & $\TopScores{5}$ & $74.29$ & $20/20$ & $0.21\%$ & $20/20$ & $0.21\%$ & - & - &  - &  - \\
    & $\TopScores{1}$ & $74.29$ & $20/20$ & $0.24\%$ & $20/20$ & $0.24\%$ &  - & - & - & -  \\ 
    &$\MemGuard$ & $74.29$ & $20/20$ & $1.57\%$ & $20/20$ & $1.65\%$ & - & - & - & -  \\ 
    \midrule 
    \multirow{7}{*}{Memorization reduction} &$\DifferentialPrivacy(\GaussianNoiseParameter=0.001)$ & $70.01$&  $20/20$ & $0.65\%$ & $20/20$ & $5.17\%$ & $20/20$ & $0.83\%$ & $20/20$ & $3.67\%$  \\  
    &$\DifferentialPrivacy(\GaussianNoiseParameter=0.002)$ & $64.11$ &  $20/20$ & $3.98\%$ & $1/20$ & $9.99\%$ & $20/20$ & $5.74\%$ & $2/20$ & $9.97\%$ \\  
    &$\DifferentialPrivacy(\GaussianNoiseParameter=0.003)$ & $59.25$ & $18/20$ & $8.14\%$ & $0/20$ & $10.00\%$ & $10/20$ & $9.34\%$ & $0/20$ & $10.00\%$ \\ 
    &$\EarlyStopping(60)$ & $73.50$ & $20/20$ & $0.25\%$ & $20/20$ & $0.28\%$ & $20/20$ & $0.51\%$ & $20/20$ & $0.63\%$ \\
    &$\EarlyStopping(40)$ & $69.15$ & $20/20$ & $0.70\%$ & $20/20$ & $3.27\%$ & $20/20$ & $1.48\%$ & $20/20$ & $3.29\%$ \\
    &$\EarlyStopping(20)$ & $67.10$ & $20/20$ & $3.18\%$ & $1/20$ & $10.00\%$ & $20/20$ & $5.68\%$ & $3/20$ & $9.51\%$ \\
    &$\AdversarialRegularization$ & $60.18$ & $20/20$ & $0.74\%$ & $20/20$ & $1.78\%$ & $20/20$ & $0.91\%$ & $20/20$ & $2.90\%$ \\
    \midrule 
    \multirow{6}{*}{Other attacks} &$\TwinsDetection$ & $74.29$ & $20/20$ & $0.20\%$ & $20/20$ & $0.20\%$ & $20/20$ & $0.64\%$ & $20/20$ & $0.74\%$ \\
    &$\NoTrainAugmentation$ & $61.59$ & $20/20$ & $0.19\%$ & $20/20$ & $0.30\%$ & $20/20$ & $0.51\%$ & $20/20$ & $0.85\%$ \\ 
    &$\GaussianNoise (\GaussianNoiseParameter=10)$ & $70.64$ & $20/20$ & $0.40\%$ & $20/20$ & $0.45\%$ & $20/20$ & $1.72\%$ & $20/20$ & $2.08\%$ \\
    &$\GaussianNoise (\GaussianNoiseParameter=20)$ & $65.97$ & $20/20$ & $1.02\%$ & $20/20$ & $1.56\%$ & $20/20$ & $5.31\%$ & $19/20$ & $7.12\%$ \\
    &$\GaussianNoise (\GaussianNoiseParameter=30)$ & $62.10$ & $20/20$ & $4.03\%$ & $20/20$ & $6.64\%$ & $16/20$ & $8.93\%$ & $6/20$ & $9.90\%$ \\ 
    &$\MarkPerturb$ & $70.49$ & $20/20$ & $0.52\%$ & $20/20$ & $0.56\%$ & $20/20$ & $3.00\%$ & $20/20$ & $3.29\%$ \\ 

    \bottomrule 
    \end{tabular}
    \vspace{1ex}
    \caption{Robustness of our proposed method on CIFAR-100 against
      countermeasures/adaptive attacks.  The most effective
      countermeasure to degrade the detection performance of our
      method is differential privacy, but it also destroyed the
      utility of the ML model.  All results were averaged over 20
      experiments.}
    \vspace{-4ex}
    \label{table:classifier:robustness}    
\end{table*}

\paragraph{Results} We summarize the robustness of our method to
these countermeasures/adaptive attacks
in~\tblref{table:classifier:robustness}.  As shown
in~\tblref{table:classifier:robustness}, masking ($\TopScores{5}$,
$\TopScores{1}$, $\MemGuard$) had limited impact on our detection
effectiveness, yielding a slightly higher
$\frac{\NumberQueriedPublished}{\TrainDatasetSize}$ but not changing
$\DetectionSuccessRate$ at all. $\DifferentialPrivacy$ and
$\EarlyStopping$ did decrease $\DetectionSuccessRate$. However, these
countermeasures damaged the utility of the trained ML model, yielding
a much smaller $\Accuracy$. Specifically, the application of
differential privacy needed a high level of privacy guarantee to
defeat our method and so added a large amount of Gaussian noise into
the training process to do so.  The added noise affected the
performance of the ML model, decreasing $\Accuracy$ to $64.11\%$, more
than $10$ percentage points lower than $\Accuracy$ with the default
training method.  Likewise, to degrade the detection performance of
our method, early stopping needed to stop the training when reaching a
small number of training epochs, at the cost of low accuracy as well,
e.g., $\Accuracy = 67.10\%$ at $20$ epochs. Among the other attacks,
detecting queried pairs and excluding data augmentation in ML model
training were not useful to counter our method. Pair detection
($\TwinsDetection$) did not work well to detect queried pairs because
we only queried the ML model with their randomly cropped versions,
which evaded pair detection. Excluding data augmentation in training
did not reduce $\DetectionSuccessRate$ but diminished the accuracy of
the ML model significantly, yielding a low $\Accuracy$ of $61.59\%$.
Adding sufficient Gaussian noise to mask the marks before training
reduced the detection effectiveness of our method but, again, it also
destroyed the utility of the ML model. For example, adding Gaussian
noise with $\GaussianNoiseParameter=30$ into marked CIFAR-100 data
reduced $\DetectionSuccessRate$ from $20/20$ to $6/20$ in condition
$\Knowledge{4}$ but also decreased $\Accuracy$ to $62.10\%$.  The last
adaptive attack, i.e., applying our marking algorithm to add
perturbations, did not decrease $\DetectionSuccessRate$ but increased
$\frac{\NumberQueriedPublished}{\TrainDatasetSize}$, at the cost of
achieving a lower $\Accuracy$ of $70.49\%$. This is because the
perturbed published data created by the marking algorithm was still
closer to the published data than to the unpublished data, which caused
the published data to appear more likely to have been used in the
training of the ML model trained on the perturbed published data.

In summary, countermeasures/adaptive attacks we considered in this
work did not defeat our auditing method or did so at the cost of
sacrificing the utility of the trained ML model; i.e., none achieved a
low $\DetectionSuccessRate$ and a high $\Accuracy$ at the same time.
Therefore, we conclude that our method achieves the \emph{robustness}
goal defined in \secref{sec:problem:design_goal} for image
classifiers.

\section{Auditing Foundation Models}
\label{sec:foundation_model}

In this section, we apply our data auditing method to detect
unauthorized use of data in foundation models.  Foundation models are
a class of large, deep neural networks for general-purpose use that
are pretrained on large-scale unlabeled data by unsupervised learning
or self-supervised learning~\cite{devlin2018:bert,
  radford2018:improving, brown2020:language,
  chen2020:simple,ramesh2021:zero, radford2021:learning,
  touvron2023:llama}.  Examples of foundation models include visual
encoders trained by self-supervised learning (e.g.,
SimCLR~\cite{chen2020:simple}), large language models (LLMs) (e.g.,
ChatGPT~\cite{radford2018:improving}), and multimodal models (e.g.,
CLIP~\cite{radford2021:learning}).  These models can be used as
backbones in various ML tasks, e.g., image
classification~\cite{deng2009:imagenet, he2016:deep}, object
detection~\cite{zhao2019:object}, sentiment
analysis~\cite{pang2002:thumbs}, text
generation~\cite{vaswani2017:attention}, and question
answering~\cite{devlin2018:bert}, by being fine-tuned on small
datasets for these tasks.

We studied the effectiveness of our proposed method on auditing
data-use in foundation models by considering three case studies: a
visual encoder trained by SimCLR~\cite{chen2020:simple},
Llama~2~\cite{touvron2023:llama}, and
CLIP~\cite{radford2021:learning}.

\subsection{Visual Encoder}  \label{sec:foundation_model:encoder}

We consider visual encoder, which is a type of foundation model used
to learn the general representations of images.  A visual encoder can
be used as a feature extractor to extract features of images in many
vision recognition tasks, e.g., image classification and object
detection. A visual encoder is an ML model that takes as input an
image and outputs its representation as a feature vector.  It is
trained by self-supervised learning (e.g.,
SimCLR~\cite{chen2020:simple}) on unlabeled data (i.e., each instance
in \TrainDataset is an image).  The loss function used by SimCLR is
Normalized Temperature-scaled Cross Entropy
(NT-Xent)~\cite{chen2020:simple}.

\subsubsection{Score Function} \label{sec:foundation_model:encoder:score_function}

We defined the score function $\ScoreFunction{\TargetModel}$ used in
the detection algorithm targeting the self-supervised visual encoder
using a black-box membership inference method introduced in
EncoderMI~\cite{liu2021:encodermi}.  The intuition behind it is that
the visual encoder $\TargetModel$ generates more similar feature
vectors of two perturbed versions of a training sample than of a
non-training sample~\cite{liu2021:encodermi}.  In other words, if
$\Data{\SampleIdx}{\PublishedTwin{\SampleIdx}}$ was used in training
$\TargetModel$ while $\Data{\SampleIdx}{\HiddenTwin{\SampleIdx}}$ was
not, then
$\CosinSim(\TargetModel(\Data{\SampleIdx}{\PublishedTwin{\SampleIdx}}[1]),
\TargetModel(\Data{\SampleIdx}{\PublishedTwin{\SampleIdx}}[2])) >
\CosinSim(\TargetModel(\Data{\SampleIdx}{\HiddenTwin{\SampleIdx}}[1]),
\TargetModel(\Data{\SampleIdx}{\HiddenTwin{\SampleIdx}}[2]))$ where
$\CosinSim$ denotes the cosine similarity,
$\Data{\SampleIdx}{\PublishedTwin{\SampleIdx}}[1]$ and
$\Data{\SampleIdx}{\PublishedTwin{\SampleIdx}}[2]$ are two perturbed
versions of $\Data{\SampleIdx}{\PublishedTwin{\SampleIdx}}$, and
$\Data{\SampleIdx}{\HiddenTwin{\SampleIdx}}[1]$ and
$\Data{\SampleIdx}{\HiddenTwin{\SampleIdx}}[2]$ are two perturbed
versions of $\Data{\SampleIdx}{\HiddenTwin{\SampleIdx}}$.  As such, we
defined the score function $\ScoreFunction{\TargetModel}$ as follows:
given an input image, we first randomly generate $\NumAugment$ of its
perturbed versions (e.g., by random cropping and flipping), and then
obtain $\NumAugment$ feature vectors using the perturbed images as
inputs to the target visual encoder; second, we compute the cosine
similarity of every pairs of feature vectors and return the sum of
cosine similarities as the score. The score function
$\ScoreFunction{\TargetModel}$ is summarized
in~\algref{alg:score_function_visual_encoder} in \appref{app:scorefunction}.

\subsubsection{Experimental Setup} \label{sec:foundation_model:encoder:setup}

\paragraph{Datasets} We used three image benchmark datasets:
CIFAR-10, CIFAR-100, and TinyImageNet, as introduced
in~\secref{sec:classifier:setup}.

\paragraph{Marking setting} We followed the setup introduced
in~\secref{sec:classifier:setup} to generate the marked dataset,
without labels needed. Our using the same marking setup
indicates that the application of our marking
algorithm is agnostic to the ML task.

\paragraph{Training setting} We followed the previous work
(e.g.,~\cite{chen2020:simple}) to train the ML model by SimCLR, which
takes as inputs a base encoder and a projection head (i.e., a
multilayer perceptron with one hidden layer).  We used ResNet18 as the
default architecture of the base encoder.  The SimCLR algorithm works
as follows: at each training step, we randomly sampled a mini-batch
(i.e., of size $512$) of images from the training set and generated
two augmented images from each sampled instance by random cropping and
resizing, random color distortion, and random Gaussian blur. The
parameters of the base encoder and the projection head were updated by
minimizing the NT-Xent loss among the generated augmented images,
i.e., maximizing the cosine similarity between any positive pair (i.e.,
two augmented images generated from the same sampled instance) and
minimizing the cosine similarity between any negative pair (i.e., two
augmented images generated from different sampled instances). We used
SGD with Nesterov Momentum~\cite{sutskever2013:importance} of $0.9$
and a weight decay of $10^{-6}$ as the optimizer, and applied a cosine
annealing schedule~\cite{loshchilov2017:sgdr} to update the learning
rate, which was set to $0.6$ initially. We trained the base encoder
and the projection head by $1{,}000$ epochs as the default, and
returned the base encoder as the visual encoder $\TargetModel$
deployed by the ML practitioner.

\paragraph{Detection setting} We applied the data-use detection algorithm to
audit the ML model.  In the detection algorithm and
the score function, we set
$\Confidence=0.025$, $\FalseDetectionRate=0.05$, and $\NumAugment=64$
as the default.

\paragraph{Metrics}
We used the following metrics for evaluation:
\begin{itemize}[nosep,leftmargin=1em,labelwidth=*,align=left]
    \item \textbf{Test accuracy of downstream classifier} (\Accuracy):
      \Accuracy is the fraction of test samples that are correctly
      classified by a downstream classifier that uses the visual
      encoder as the backbone and is fine-tuned on a small set of
      data.  We followed previous work (e.g.,~\cite{chen2020:simple})
      to fine-tune the downstream classifier on $10\%$ of the clean
      training samples with their labels (i.e., $5{,}000$ clean
      CIFAR-10 data, $5{,}000$ clean CIFAR-100 data or $10{,}000$
      clean TinyImageNet data).  A higher \Accuracy indicates a better
      performance of the visual encoder.
    \item \textbf{Detection success rate} ($\DetectionSuccessRate$):
      please see the description of this metric in
      \secref{sec:classifier:setup}.
    \item \textbf{Ratio between the number of queried published data
      and the total number of training samples}
      ($\frac{\NumberQueriedPublished}{\TrainDatasetSize}$):
      please see the description of this metric in
      \secref{sec:classifier:setup}.
\end{itemize}

\subsubsection{Experimental Results} \label{sec:foundation_model:encoder:result}

The overall experimental results on three visual benchmarks are
presented in \tblref{table:visual_encoder}.  As shown in
\tblref{table:visual_encoder}, our proposed method achieved highly
effective detection performance on auditing data in visual encoders,
yielding a $19/20$ $\DetectionSuccessRate$ for CIFAR-10 and a $20/20$
$\DetectionSuccessRate$ for CIFAR-100 and TinyImageNet.

\begin{table}[ht!]
    \centering
    \begin{tabular}{@{}lr@{\hspace{0.5em}}rr@{\hspace{0.5em}}r@{}}
    \toprule
    \multicolumn{1}{c}{}& \multicolumn{1}{c}{$\DetectionSuccessRate$} & \multicolumn{1}{c}{$\frac{\NumberQueriedPublished}{\TrainDatasetSize}$}\\ 
    \midrule 
    CIFAR-10& $19/20$ & $7.12\%$ \\ 
    CIFAR-100& $20/20$ & $7.28\%$ \\ 
    TinyImageNet& $20/20$ & $7.82\%$ \\ 
    \bottomrule 
    \end{tabular} 
    \vspace{1ex} 
    \caption{Results on auditing data in visual encoder trained by
      SimCLR, under an upper bound of $\FalseDetectionRate = 0.05$ on
      the false-detection rate. $10\%$ of training samples were
      marked. All results were averaged over 20 experiments.}
    \vspace{-3ex} 
    \label{table:visual_encoder}
\end{table}

\begin{figure}
    \centering
    \begin{subfigure}[b]{0.23\textwidth}
        \setlength\figureheight{1.9in}
        \centering
        \resizebox{\textwidth}{!}{\begin{tikzpicture}[font=\huge]

    \definecolor{crimson2143940}{RGB}{214,39,40}
    \definecolor{darkgray176}{RGB}{176,176,176}
    \definecolor{darkorange25512714}{RGB}{255,127,14}
    \definecolor{forestgreen4416044}{RGB}{44,160,44}
    \definecolor{lightgray204}{RGB}{204,204,204}
    \definecolor{steelblue31119180}{RGB}{31,119,180}
    
    \begin{axis}[
        legend cell align={left},
        legend style={
            fill opacity=0.8,
            draw opacity=1,
            text opacity=1,
            at={(0.97,0.03)},
            anchor=south east,
            draw=lightgray204
        },
        tick align=outside,
        tick pos=left,
        x grid style={darkgray176},
        xlabel={Epochs},
        xmin=100, xmax=1100,
        xtick style={color=black},
        y grid style={darkgray176},
        ylabel={$\DetectionSuccessRate$},
        ymin=-0.05, ymax=1.05,
        ytick style={color=black}
    ]
    \addplot [line width=3, steelblue31119180, mark=*, mark size=5, mark options={solid}]
    table {%
    200 0.1
    400 0.4
    600 0.9
    800 1
    1000 1
    };
    \end{axis}
    
    \end{tikzpicture}}
        \caption{Detection performance}
      \end{subfigure}
      \hfill
      \begin{subfigure}[b]{0.235\textwidth}
        \setlength\figureheight{1.9in}
        \centering
        \resizebox{\textwidth}{!}{
\begin{tikzpicture}[font=\huge]

    \definecolor{crimson2143940}{RGB}{214,39,40}
    \definecolor{darkgray176}{RGB}{176,176,176}
    \definecolor{darkorange25512714}{RGB}{255,127,14}
    \definecolor{forestgreen4416044}{RGB}{44,160,44}
    \definecolor{lightgray204}{RGB}{204,204,204}
    \definecolor{steelblue31119180}{RGB}{31,119,180}
    
    \begin{axis}[
        legend cell align={left},
        legend style={
            fill opacity=0.8,
            draw opacity=1,
            text opacity=1,
            at={(0.97,0.03)},
            anchor=south east,
            draw=lightgray204
        },
        tick align=outside,
        tick pos=left,
        x grid style={darkgray176},
        xlabel={Epochs},
        xmin=100, xmax=1100,
        xtick style={color=black},
        y grid style={darkgray176},
        ylabel={$\Accuracy \%$},
        ymin=70.10, ymax=71.33,
        ytick style={color=black}
    ]
    \addplot [line width=3, darkorange25512714, mark=*, mark size=5, mark options={solid}]
table {%
200 70.21
400 70.71
600 71.01
800 71.25
1000 71.28
};
\end{axis}

\end{tikzpicture}}
        \caption{Encoder utility}
      \end{subfigure}
      \vspace{-2ex}
    \caption{The impact of epochs on the detection performance and encoder utility. 
    The evaluated encoder was trained by SimCLR on marked CIFAR-100 ($10\%$ are marked). The results are averaged over 20 experiments.}
    \vspace{-2ex}
    \label{fig:impact_epochs_encoder}
\end{figure}
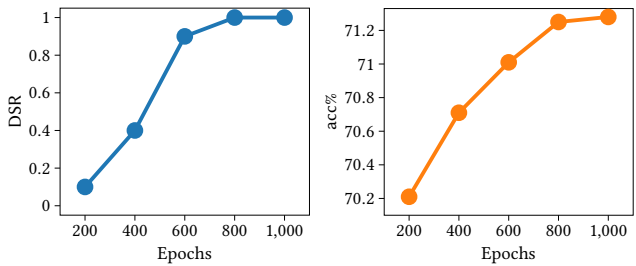

We investigated the impact of training epochs of visual encoder on the
detection performance of the auditing method.  We trained visual
encoders on marked CIFAR-100 by epochs of $200$, $400$, $600$, $800$,
and $1{,}000$ ($1{,}000$ is the default number of epochs). As shown in
\figref{fig:impact_epochs_encoder}, when we trained the encoder with a
smaller number of epochs, the encoder memorized the training samples
less and thus we had a lower $\DetectionSuccessRate$.  However,
training encoder with fewer epochs yielded modestly lower
encoder utility, measured by the test accuracy of the downstream
classifier (i.e., $\Accuracy$).  This suggests that early stopping
(i.e., training with a small number of epochs) can degrade the
detection performance of our method, but cannot completely alleviate
the trade-off between evading detection and encoder utility.

\subsection{Llama 2}  \label{sec:foundation_model:llama}

In this section, we study the application of data auditing to a large
language model (LLM).  An LLM is a type of large ML model that can
understand and generate human language.  Here we consider
Llama~2~\cite{touvron2023:llama} published by Meta AI in 2023, which
is an open-sourced LLM with notable performance and, more importantly,
is free for research~\cite{touvron2023:llama}.  Specifically, Llama~2
is a family of autoregressive models that generate text by predicting
the next token based on the previous ones.  They are designed with a
transformer architecture~\cite{vaswani2017:attention} with parameters
ranging from 7 billion to 70 billion, and pretrained and fine-tuned on
massive text datasets containing trillions of tokens collected from
public sources~\cite{touvron2023:llama}.  Considering Llama~2 as the
ML model \TargetModel in \eqnref{eq:training}, each instance in the
training dataset or fine-tuning dataset is a sequence of tokens (e.g.,
by a tokenizer defining a token vocabulary $\TokenVocabulary$) of
length $\SequenceLength$, i.e., $\TrainSample{\SampleIdx} =
\Token{\SampleIdx}{1}\Token{\SampleIdx}{2}\dots\Token{\SampleIdx}{\SequenceLength}$
($\Token{\SampleIdx}{\TokenIdx} \in \TokenVocabulary$ for any
$\TokenIdx \in \{1,2,\dots,\SequenceLength\}$) and the loss function
is defined as:
\begin{equation}  \label{eq:loss_text}
  \LossFunction(\TargetModel, \TrainSample{\SampleIdx}) = \sum_{\TokenIdx=1}^{\SequenceLength} -\log \arrComponent{\TargetModel(\Token{\SampleIdx}{1}\dots\Token{\SampleIdx}{\TokenIdx-1})}{\Token{\SampleIdx}{\TokenIdx}},
\end{equation}
where $\arrComponent{\TargetModel(\Token{\SampleIdx}{1}\dots\Token{\SampleIdx}{\TokenIdx-1})}{\Token{\SampleIdx}{\TokenIdx}}$ denotes the $\Token{\SampleIdx}{\TokenIdx}$-th component of vector $\TargetModel(\Token{\SampleIdx}{1}\dots\Token{\SampleIdx}{\TokenIdx-1})$.

It is challenging to conduct lab-level experiments on auditing data in
a pretrained Llama~2 because pretraining Llama~2 on a massive text
corpus needs a huge amount of computing resources.  Therefore, instead
of applying our data auditing method to the pretrained Llama 2, we
mainly focus on a Llama~2 fine-tuning setting.

\subsubsection{Score Function} \label{sec:foundation_model:llama:score_function}
We used the negative loss, a simple and effective membership inference
metric~\cite{carlini2021:extracting}, as the score function. Formally,
given a text sample $\Data{\SampleIdx}{\genericInd}$, we have
$\ScoreFunction{\TargetModel}(\Data{\SampleIdx}{\genericInd}) =
-\LossFunction(\TargetModel, \Data{\SampleIdx}{\genericInd})$, where
$\LossFunction(\TargetModel, \Data{\SampleIdx}{\genericInd})$ is
defined in \eqnref{eq:loss_text}.

\subsubsection{Experimental Setup} \label{sec:foundation_model:llama:setup}

\paragraph{Datasets}
We used three text datasets: SST2~\cite{socher2013:recursive}, AG's news~\cite{zhang2015:character}, and TweetEval (emoji)~\cite{mohammad2018:semeval}:
\begin{itemize}[nosep,leftmargin=1em,labelwidth=*,align=left]
  \item \textbf{SST2}: SST2 is a dataset containing sentences used for sentiment analysis (i.e., there are $2$ classes, ``Negative'' and ``Positive'').
      In SST2, there are $67{,}300$ training samples and $872$ validation samples that we used for testing.
  \item \textbf{AG's news}: AG's news is a dataset containing sentences partitioned into $4$ classes, ``World'', ``Sports'', ``Business'', and ``Sci/Tech''.
      In AG's news, there are $120{,}000$ training samples and $7{,}600$ test samples.
  \item \textbf{TweetEval (emoji)}: TweetEval (emoji) is a dataset containing sentences partitioned into $20$ classes. In TweetEval (emoji),
      there are $100{,}000$ training samples and $50{,}000$ test samples.
\end{itemize}

\paragraph{Marking setting} In each experiment, we uniformly at
random sampled a subset of training samples of a dataset as
$\TrainingSamples$ (e.g., $\setSize{\TrainingSamples}=10{,}000$).
From $\TrainingSamples$, we uniformly at random sampled
$\totalSamples=1{,}000$ sentences
$\{\Data{\SampleIdx}\}_{\SampleIdx=1}^{\totalSamples}$ assumed to be
owned by a data owner. We applied our data marking algorithm to generate the
published data
$\{\Data{\SampleIdx}{\PublishedTwin{\SampleIdx}}\}_{\SampleIdx =
  1}^{\totalSamples}$ and the unpublished data
$\{\Data{\SampleIdx}{\HiddenTwin{\SampleIdx}}\}_{\SampleIdx =
  1}^{\totalSamples}$ for
$\{\Data{\SampleIdx}\}_{\SampleIdx=1}^{\totalSamples}$.  In
\eqnref{eq:twins_data}, we defined the
distance function by Levenshtein
distance~\cite{levenshtein1966:binary} and the utility difference
function by semantic dissimilarity~\cite{devlin2018:bert}.  Instead of
solving \eqnref{eq:twins_data} exactly, we approximated it by using a
paraphraser model (e.g.,~\cite{chatgpt_paraphraser}) to generate two
semantically similar but distinct sentences.  We set
$\PublishedInTrainDataset =
\{\Data{\SampleIdx}{\PublishedTwin{\SampleIdx}}\}_{\SampleIdx =
  1}^{\totalSamples}$.  As such, we constituted the training dataset
collected by the ML practitioner as $\TrainDataset =(\TrainingSamples
\setminus \{\Data{\SampleIdx}\}_{\SampleIdx=1}^{\totalSamples}) \cup
\PublishedInTrainDataset$, labeled correctly (i.e., using their
original labels).

\paragraph{Fine-tuning setting}
We used \texttt{Llama-2-7b-chat-hf} Llama 2 model released in Hugging
Face\footnote{https://huggingface.co/meta-llama/Llama-2-7b-chat-hf} as
the base model.  We used QLoRA~\cite{dettmers2024:qlora} to fine-tune
the Llama~2 model on the marked dataset, where we applied
AdamW~\cite{loshchilov2018:fixing} as the optimizer that was also used
to pretrain Llama~2 by Meta AI~\cite{touvron2023:llama}. We fine-tuned
the model with a learning rate of $2\times10^{-4}$. The fine-tuned
Llama~2 is the ML model deployed by the ML practitioner.

\paragraph{Detection setting} We applied the detection algorithm to
detect a given ML model using a set of pairs of generated published
data and unpublished data. In the detection algorithm, we set
$\Confidence=0.025$ and $\FalseDetectionRate=0.05$.

\paragraph{Metrics}
We used the following metrics to evaluate methods:
\begin{itemize}[nosep,leftmargin=1em,labelwidth=*,align=left]
  \item \textbf{Test accuracy} (\Accuracy): \Accuracy is the fraction
    of test samples that were correctly classified by the fine-tuned
    Llama~2. A higher \Accuracy indicates a better performance of the
    fine-tuned model.
  \item \textbf{Detection success rate} ($\DetectionSuccessRate$):
    please see the description of this metric in
    \secref{sec:classifier:setup}.
  \item \textbf{Ratio between the number of queried published data and
    the total number of training samples}
    ($\frac{\NumberQueriedPublished}{\TrainDatasetSize}$):
    please see the description of this metric in
    \secref{sec:classifier:setup}.
\end{itemize}

\subsubsection{Experimental Results} \label{sec:foundation_model:llama:result}
The results on applying our auditing method to the fine-tuned Llama~2
on three marked datasets are presented in
\tblref{table:llama2:overview}. As shown in
\tblref{table:llama2:overview}, when we tested the detection method on
the pretrained Llama~2 (i.e., in the row of ``Epoch 0''), we obtained
a $\DetectionSuccessRate$ of $0/20$, indicating that the Llama~2 is
not pretrained on the published data. If true, this result
empirically confirms the bounded false-detection rate of our method.
When we fine-tuned Llama~2 on the marked datasets by only 1 epoch, the
accuracy of the fine-tuned Llama~2 model increased from $63.07\%$ to
$95.33\%$ for SST2, from $28.41\%$ to $91.69\%$ for AG's news, and
from $16.58\%$ to $40.49\%$ for TweetEval. At the same time, our
method achieved a $\DetectionSuccessRate$ of $20/20$ on the fine-tuned
model, which demonstrates the effectiveness of our method.
Fine-tuning Llama~2 with more epochs increased the accuracy slightly
(e.g., from $95.33\%$ to $95.56\%$ for SST2, from $91.69\%$ to
$92.33\%$ for AG's news, and from $40.49\%$ to $43.03\%$ for
TweetEval) but leads to a much lower
$\frac{\NumberQueriedPublished}{\TrainDatasetSize}$.  This is
because fine-tuning the model for more epochs memorizes the
fine-tuning samples more and the detection method needs fewer
queries to the model to detect their use.

\begin{table*}[h!]
  \centering
  \begin{minipage}{0.63\textwidth}
  \vspace{-5ex} 
  \centering
  \begin{tabular}{@{}lr@{\hspace{0.5em}}rr@{\hspace{0.5em}}rr@{\hspace{0.5em}}rr@{\hspace{0.5em}}rr@{\hspace{0.5em}}r@{}}
    \toprule
    \multicolumn{1}{c}{} & \multicolumn{3}{c}{SST2}& \multicolumn{3}{c}{AG's news} & \multicolumn{3}{c}{TweetEval (emoji)}\\ 
    & \multicolumn{1}{c}{$\Accuracy\%$} & \multicolumn{1}{c}{$\DetectionSuccessRate$} & \multicolumn{1}{c}{$\frac{\NumberQueriedPublished}{\TrainDatasetSize}$} & \multicolumn{1}{c}{$\Accuracy\%$} & \multicolumn{1}{c}{$\DetectionSuccessRate$} & \multicolumn{1}{c}{$\frac{\NumberQueriedPublished}{\TrainDatasetSize}$}  & \multicolumn{1}{c}{$\Accuracy\%$} & \multicolumn{1}{c}{$\DetectionSuccessRate$} & \multicolumn{1}{c}{$\frac{\NumberQueriedPublished}{\TrainDatasetSize}$} \\ 
    \midrule 
    Epoch 0 & $63.07$ & $0/20$ & $10.00\%$ & $28.41$ & $0/20$ & $10.00\%$ & $16.58$ & $0/20$ & $10.00\%$ \\
    Epoch 1 & $95.33$ & $20/20$ & $2.87\%$ & $91.69$ & $20/20$ & $2.97\%$ & $40.49$ & $20/20$ & $3.89\%$\\
    Epoch 2 & $95.26$ & $20/20$ & $0.22\%$ & $91.68$ & $20/20$ & $0.23\%$ & $41.88$ & $20/20$ & $0.26\%$\\
    Epoch 3 & $95.56$ & $20/20$ & $0.12\%$ & $92.33$ & $20/20$ & $0.12\%$ & $43.03$ & $20/20$ & $0.12\%$\\
    \bottomrule 
    \end{tabular}
    \vspace{1ex}
    \caption{Overall performance of our proposed method on Llama~2
      fine-tuned on marked text datasets ($10\%$ of fine-tuning
      samples were marked) for different numbers of epochs, under an
      upper bound of $\FalseDetectionRate = 0.05$ on the
      false-detection rate.  All results were averaged over 20
      experiments.}
    \vspace{-4ex} 
    \label{table:llama2:overview} 
  \end{minipage}%
  \hfill
  \begin{minipage}{0.33\textwidth}
  \centering
  \begin{tabular}{@{}lr@{\hspace{0.5em}}rr@{\hspace{0.5em}}rr@{\hspace{0.5em}}rr@{\hspace{0.5em}}rr@{\hspace{0.5em}}rr@{\hspace{0.5em}}rr@{\hspace{0.5em}}rr@{\hspace{0.5em}}r@{}}
    \toprule
    & \multicolumn{1}{c}{$\Accuracy\%$} & \multicolumn{1}{c}{$\DetectionSuccessRate$} & \multicolumn{1}{c}{$\frac{\NumberQueriedPublished}{\TrainDatasetSize}$} \\ 
    \midrule 
    Epoch 0 & $80.73$ & $0/20$ & $10.00\%$ \\
    Epoch 1 & $88.44$ & $20/20$ & $6.99\%$ \\
    Epoch 2 & $88.53$ & $20/20$ & $2.31\%$ \\
    Epoch 3 & $88.53$ & $20/20$ & $1.21\%$ \\
    \bottomrule 
    \end{tabular}
    \vspace{1ex}
    \caption{Overall performance of our proposed method on CLIP
      fine-tuned on marked Flickr30k ($10\%$ of fine-tuning samples
      were marked) for different numbers of epochs, under an upper
      bound of $\FalseDetectionRate = 0.05$ on the false-detection
      rate.  All results were averaged over 20 experiments.}
    \vspace{-4ex} 
    \label{table:clip:overview}    
  \end{minipage}
  \end{table*}

\subsection{CLIP}  \label{sec:foundation_model:clip}

In this section, we apply data auditing to a multimodal
model~\cite{radford2021:learning, ramesh2021:zero}.  A multimodal model
is a type of ML model that can understand and process various types of
data, e.g., image, text, and audio.  We considered Contrastive
Language-Image Pretraining (CLIP)~\cite{radford2021:learning},
developed by OpenAI in 2021, as our study case.  CLIP is a
vision-language model consisting of a visual encoder and a text
encoder used to extract the features of the input image and text,
respectively.  It takes as inputs an image and a text and returns
their corresponding feature vectors.  CLIP is known for its notable
performance in image-text similarity and zero-shot image
classification~\cite{radford2021:learning}.  As in
\eqnref{eq:training}, each instance is an image with its caption
(i.e., a pair of image and text) and the loss function is the cross
entropy loss used to push matched images and texts closer in the
shared latent space while pushing unrelated pairs apart.

The CLIP model released by OpenAI was pretrained on 400 million image
and text pairs collected from the
Internet~\cite{radford2021:learning}.  While it is challenging to
pretrain such a large model on a huge number of pairs using lab-level
computing resources, we aim to fine-tune the CLIP on a small (marked)
dataset and test our auditing method on the fine-tuned CLIP.

\subsubsection{Score Function} \label{sec:foundation_model:clip:score_function}

We defined the score function $\ScoreFunction{\TargetModel}$ by a
recently proposed membership inference on
CLIP~\cite{ko2023:practical}.  It uses cosine similarity between the
two feature vectors returned by the CLIP model as the inference
metric~\cite{ko2023:practical}. Formally, given an image-text sample
$\Data{\SampleIdx}{\genericInd} =
(\Image{\SampleIdx}{\genericInd}, \Caption{\SampleIdx}{\genericInd})$,
we have $\ScoreFunction{\TargetModel}(\Data{\SampleIdx}{\genericInd})
= \CosinSim(\VisualEncoder(\Image{\SampleIdx}{\genericInd}),
\TextEncoder(\Caption{\SampleIdx}{\genericInd}))$, where
$\VisualEncoder$ and $\TextEncoder$ are the visual encoder and text
encoder of $\TargetModel$, and $\CosinSim$ denotes cosine similarity.

\subsubsection{Experimental Setup} \label{sec:foundation_model:clip:setup}

\paragraph{Datasets} We used the Flickr30k~\cite{young2014:image}
dataset, which contains more than $31{,}000$ images with captions.  We
used the first $25{,}000$ as training samples and the remaining as
test samples.

\paragraph{Marking setting} In each experiment, we uniformly at
random sampled $\totalSamples$ images with captions from training
samples $\TrainingSamples$.  We set
$\frac{\totalSamples}{\setSize{\TrainingSamples}} = 10\%$, i.e.,
$\totalSamples=2{,}500$. We assumed these $\totalSamples$ captioned
images are owned by a data owner. We applied our data marking algorithm to
generate the published data
$\{\Data{\SampleIdx}{\PublishedTwin{\SampleIdx}}\}_{\SampleIdx =
  1}^{\totalSamples}$ and the unpublished data
$\{\Data{\SampleIdx}{\HiddenTwin{\SampleIdx}}\}_{\SampleIdx =
  1}^{\totalSamples}$ for
$\{\Data{\SampleIdx}\}_{\SampleIdx=1}^{\totalSamples}$.
In the marking algorithm, given a raw datum (i.e., an image with its
caption), we followed the marking setting in
\secref{sec:classifier:setup} to generate two marked images and then
randomly sampled one with its original caption as the published data,
keeping the other as the unpublished data.
We set $\PublishedInTrainDataset =
\{\Data{\SampleIdx}{\PublishedTwin{\SampleIdx}}\}_{\SampleIdx =
  1}^{\totalSamples}$.  As such, we constituted the training dataset
collected by the ML practitioner as $\TrainDataset =(\TrainingSamples
\setminus \{\Data{\SampleIdx}\}_{\SampleIdx=1}^{\totalSamples}) \cup
\PublishedInTrainDataset$.

\paragraph{Fine-tuning setting} We used the CLIP model released by
OpenAI\footnote{https://github.com/openai/CLIP} as the base model.  We
fine-tuned the CLIP model on the marked dataset $\TrainDataset$,
following the pretraining algorithm used by
OpenAI~\cite{radford2021:learning}.  We used a batch size of $256$ and
applied Adam~\cite{kingma2014:adam} with a learning rate of $10^{-5}$
as the optimizer.  The fine-tuned CLIP including the visual encoder
and text encoder is the ML model deployed by the ML practitioner.

\paragraph{Detection setting}
We applied our data-use detection algorithm to detect a given ML model using a
set of pairs of generated published data and unpublished data. In
the detection algorithm, we set $\Confidence=0.025$ and
$\FalseDetectionRate=0.05$.

\paragraph{Metrics}
We used the following metrics for evaluation:
\begin{itemize}[nosep,leftmargin=1em,labelwidth=*,align=left]
    \item \textbf{Test accuracy} (\Accuracy): We randomly divided the test samples
      into batches (each is $256$ at most). For each batch, we measured the fraction 
      of texts correctly matched to images and the fraction of images correctly
      matched to texts, by the (fine-tuned) CLIP model. We used 
      the fraction of correct matching averaged over batches as the test accuracy \Accuracy.
    \item \textbf{Detection success rate} ($\DetectionSuccessRate$):
      please see the description of this metric in
      \secref{sec:classifier:setup}.
    \item \textbf{Ratio between the number of queried published data
      and the total number of training samples}
      ($\frac{\NumberQueriedPublished}{\TrainDatasetSize}$):
      please see the description of this metric in
      \secref{sec:classifier:setup}.
\end{itemize}

\subsubsection{Experimental Results} \label{sec:foundation_model:clip:result}

The overall performance of our data auditing method applied in
fine-tuned CLIP is presented in \tblref{table:clip:overview}. As shown
in \tblref{table:clip:overview}, when we audited the CLIP model
released by OpenAI, we obtained a $0/20$ $\DetectionSuccessRate$,
which indicates that the pretrained CLIP model was not trained on our
published data.  If it is true that the CLIP model is not, this result
empirically confirms the upper bound on the false-detection rate of
our method.  When we fine-tuned the CLIP model by the marked Flickr30k
dataset, $\Accuracy$ increased from $80.73\%$ to $88.44\%$ while
$\DetectionSuccessRate$ increased to $20/20$, which demonstrates that
our method is highly effective to detect the use of published data in
the fine-tuned CLIP even when it is fine-tuned by only 1 epoch.  When
we fine-tuned the model for more epochs (e.g, $3$ epochs), $\Accuracy$
did not significantly increase.  With more fine-tuning epochs, we
still got a $\DetectionSuccessRate$ of $20/20$ but a smaller
$\frac{\NumberQueriedPublished}{\TrainDatasetSize}$.  Fine-tuning by
more epochs made the model memorize the fine-tuning samples more and
thus we needed fewer queries to the model in the detection step.

\section{Discussion and Limitations} \label{sec:discussion}

\subsection{Minimal Number of Marked Data Required in Auditing}  \label{sec:discussion:minimal_number}

The minimal number of marked (published) data for which our method can
detect its use depends on two factors: the memorization of training
data by the ML model and the effectiveness of (contrastive) membership
inference.  For example, as shown in \secref{sec:classifier:results},
the CIFAR-100 and TinyImagenet classifiers memorized their training
samples more than the CIFAR-10 classifier, and so the data owner
needed much less marked data to audit for data use in the CIFAR-100
and TinyImagenet classifiers than in the CIFAR-10 classifier.  The
effectiveness of (contrastive) membership inference also affects the
minimal number of data items for which our method can detect use,
i.e., a stronger membership inference method will allow our method to
detect the use of fewer data.  Therefore, we believe that any
developed stronger membership inference methods in the future will
benefit our technique.

\subsection{Adaptive Attacks to Data Auditing Applied in Foundation Models} \label{sec:discussion:adaptive_attack}

Once the ML practitioner realizes that the data auditing is being
applied, he might utilize adaptive attacks aiming to defeat the
auditing method when training his foundation models.  Some adaptive
attacks we considered for the image classifier (see
\secref{sec:classifier:results:adaptive_attack}) like early stopping,
regularization, and differential privacy, can be used to mitigate the
memorization of training/fine-tuning samples of foundation
models. Therefore, these adaptive attacks could degrade the
effectiveness or efficiency of our detection method. In addition,
there are some methods used to mitigate membership inference in LLMs,
e.g., model parameter quantization/rounding~\cite{pan2020:privacy}.
Any defense against membership inference in foundation models can be
used as an adaptive attack. However, the application of these adaptive
attacks will decrease the utility of the foundation
models~\cite{pan2020:privacy}.

Since developers of foundation models usually aim to develop a
powerful foundation model, they might hesitate to apply these adaptive
methods since they will lose some model utility. As such, our data
auditing method can pressure those developers of large foundation
models to seek data-use authorization from the data owners before
using their data.

\subsection{Cost of Experiments on Foundation Models}
\label{sec:discussion:limitations}

In our experiments on auditing data use in foundation models (e.g.,
Llama~2 in \secref{sec:foundation_model:llama} and CLIP in
\secref{sec:foundation_model:clip}), we only considered model
fine-tuning due to our limited computing resources. From the results
shown in \secref{sec:foundation_model:llama:result} and
\secref{sec:foundation_model:clip:result}, our proposed method
achieves good performance on detecting the use of data in fine-tuning
Llama~2 and CLIP. We do believe that the effective detection
performance of our method can be generalized to other types of
foundation models and the settings where we audit the use of data in
pretrained foundation models. This is because large foundation models
memorize their training samples and thus are vulnerable to membership inference and
other privacy attacks, as shown by existing works
(e.g.,~\cite{carlini2021:extracting,pan2020:privacy,liu2021:encodermi,ko2023:practical,shi2024:detecting}).

\subsection{Toward Verifiable Machine Unlearning}  \label{sec:discussion:machine_unlearning}

One direct application of our data-auditing method is to verify
machine unlearning. Machine unlearning is a class of methods that
enable an ML model to forget some of its training samples upon the
request of their owners. While there are recent efforts to develop
machine unlearning algorithms~\cite{bourtoule2021:machine}, few focus
on the verification of machine unlearning, i.e., verifying if the
requested data has indeed been forgotten by the target
model~\cite{sommer2022:towards}. Our proposed method can be a good fit
for verifying machine unlearning. Specifically, each data owner
utilizes our marking algorithm to generate
published data and hidden data.  Upon the approval of data owners, a
ML practitioner collects their published data and trains an ML model
that can be verified by the data owner using our detection algorithm.  
If a data owner sends a request to the ML
practitioner to delete her data from the ML model, the ML practitioner
will utilize a machine unlearning algorithm to remove her data from
his ML model and then inform the data owner of the successful
removal. The data owner can utilize the detection algorithm
to verify if the updated ML model still uses
her published data. Our results in
\secref{sec:classifier:results:overall} show that our auditing method
remains highly effective even when multiple data owners audit their
data independently.

\subsection{Proving a Claim of Data Use}  \label{sec:discussion:deployment}

Though our technique enables a data owner to determine whether an ML
practitioner used her data without authorization, it alone does not
suffice to enable the data owner to convince a third party.  To
convince a third party, the data owner should commit to
$\{\Data{\SampleIdx}{\PublishedTwin{\SampleIdx}}\}_{\SampleIdx=1}^{\totalSamples}$
and
$\{\Data{\SampleIdx}{\HiddenTwin{\SampleIdx}}\}_{\SampleIdx=1}^{\totalSamples}$
prior to publishing the former, e.g., by escrowing a cryptographic
commitment to these data with the third party.  Upon detecting use of
her data by an ML practitioner, the data owner can open these
commitments to enable the third party perform our hypothesis test on
the ML model itself, for example.  To enable a third party to
replicate the data-owner's test result exactly, the data owner could
provide the seed to a random number generator to drive the sequence of
selections (WoR) from $\{\indicatorVal{1}, \ldots,
\indicatorVal{\totalSamples}\}$ in the test (see
\secref{sec:proposed_framework:detection:wor}).  However, to protect
an ML practitioner from being framed by a malicious data owner, the
data owner should be unable to freely choose this seed; e.g., it could
be set to be a cryptographic hash of the commitments to
$\{\Data{\SampleIdx}{\PublishedTwin{\SampleIdx}}\}_{\SampleIdx=1}^{\totalSamples}$
and
$\{\Data{\SampleIdx}{\HiddenTwin{\SampleIdx}}\}_{\SampleIdx=1}^{\totalSamples}$.

\section{Conclusion} \label{sec:conclusion}

In this paper, we proposed a general framework allowing a data owner
to audit ML models for the use of her data.  Our data auditing
framework leverages any membership-inference technique, folding it
into a sequential hypothesis test for which we can quantify the
false-detection rate.  Through evaluations of our proposed framework
in the cases of an image classifier and various foundation models, we
showed that it is effective, robust, and general across different
types of ML models and settings. We thus believe our proposed
framework provides a useful tool for data owners to audit ML models
for the use of their data.

\section*{Acknowledgments}

We thank the anonymous reviewers for their comments. 
This work was supported in part by NSF grants 2112562, 2125977, 
1937787, and 2131859, as well as ARO grant No. W911NF2110182. 

\bibliographystyle{ACM-Reference-Format}
\bibliography{full, references}

\appendix


\section{Data Marking Algorithm and Data-Use Detection Algorithm} \label{app:algorithm}

\paragraph{Data marking algorithm}
The marking algorithm introduced in \secref{sec:proposed_framework:marking} is summarized in~\algref{alg:marking}. 
The marked data generation step creates a pair $(\Data{\SampleIdx}{0},
\Data{\SampleIdx}{1})$, both crafted from the raw datum
\Data{\SampleIdx}.  Taking the example where \Data{\SampleIdx} is an
image, we set $\Data{\SampleIdx}{0} \gets \Data{\SampleIdx} +
\Mark{\SampleIdx}$ and $\Data{\SampleIdx}{1} \gets \Data{\SampleIdx} -
\Mark{\SampleIdx}$ where $\Mark{\SampleIdx}$ is the added mark.  The
random sampling step selects $\PublishedTwin{\SampleIdx} \getsr \{0,
1\}$ and publishes \Data{\SampleIdx}{\PublishedTwin{\SampleIdx}},
keeping \Data{\SampleIdx}{\HiddenTwin{\SampleIdx}} secret.

\begin{algorithm}[!t]
  \caption{Data marking algorithm}  \label{alg:marking}
  \begin{algorithmic}[1]
  \REQUIRE A bound $\MarkBound \in \realsPos$ and a set
  $\{\Data{\SampleIdx}\}_{\SampleIdx=1}^{\totalSamples}$ of raw
  data.
  
  \FOR{$\SampleIdx \gets 1, 2, \dots, \totalSamples$}
  \STATE \texttt{//Step I: Marked data generation}
  \STATE Generate $(\Data{\SampleIdx}{0}, \Data{\SampleIdx}{1})$ by solving \eqnref{eq:twins_data}.
  
  \STATE \texttt{//Step II: Random sampling}
  \STATE $\PublishedTwin{\SampleIdx} \getsr \{0, 1\}$. 
  \ENDFOR
  
  \ENSURE
  $\{\Data{\SampleIdx}{\PublishedTwin{\SampleIdx}}\}_{\SampleIdx=1}^{\totalSamples}$
  to publish and
  $\{\Data{\SampleIdx}{\HiddenTwin{\SampleIdx}}\}_{\SampleIdx=1}^{\totalSamples}$
  to hide.
  \end{algorithmic}
\end{algorithm}

\paragraph{Data-use detection algorithm}
Our detection algorithm introduced in \secref{sec:proposed_framework:detection} 
is summarized in \algref{alg:detection}.  At each
time step, the data owner samples an $\SampleIdx \in \{1, \ldots,
\totalSamples\}$ uniformly at random WoR and estimates $\TotalSuccess$
based on the currently obtained measurements using a
prior-posterior-ratio martingale (PPRM)~\cite{waudby2020:confidence} that takes as inputs the
sequence of measurements so far, the size of the population
$\totalSamples$, and the confidence level $\Confidence$.  It returns a
confidence interval for $\TotalSuccess$.  If the interval (i.e., its
lower bound) is equal to or larger than a preselected threshold
$\sumThreshold$, the data owner stops sampling and rejects the null
hypothesis; otherwise, she continues the sampling.

\begin{algorithm}[!t]
  \caption{Data-use detection algorithm}  \label{alg:detection}
  \begin{algorithmic}[1]
  \REQUIRE Published data
  $\{\Data{\SampleIdx}{\PublishedTwin{\SampleIdx}}\}_{\SampleIdx =
    1}^{\totalSamples}$, hidden data
  $\{\Data{\SampleIdx}{\HiddenTwin{\SampleIdx}}\}_{\SampleIdx =
    1}^{\totalSamples}$, confidence level $\Confidence$ of PPRM, a
  desired upper bound \FalseDetectionRate ($\FalseDetectionRate >
  \Confidence$) on the false-detection rate, and a score function
  \ScoreFunction{\TargetModel}.
  \STATE Initialize $\DetectionResult \gets \mbox{False}$;
  \STATE Initialize the measurement sequence $\MeasurementSequence \gets \emptyset$;
  \STATE Find a min $\sumThreshold \in \{\ceil{\frac{\totalSamples}{2}}, \ldots, \totalSamples\}$ s.t. $\bigg(\frac{\exp\big(\frac{2\sumThreshold}{\totalSamples} - 1\big)}{\big(\frac{2\sumThreshold}{\totalSamples}\big)^{\frac{2\sumThreshold}{\totalSamples}}}\bigg)^{\frac{\totalSamples}{2}} \leq \FalseDetectionRate - \Confidence$;
  \FOR{$\ObservationTime \gets 1, 2, \dots, \totalSamples$}
  \STATE Sample an $\SampleIdx$ uniformly at random WoR from $\{1, 2, \ldots, \totalSamples\}$;
  \STATE \texttt{//Append the new measurement to the sequence.}
  \STATE $\MeasurementSequence \gets \MeasurementSequence \cup \{\IndicatorFunction{\ScoreFunction{\TargetModel}(\Data{\SampleIdx}{\PublishedTwin{\SampleIdx}})>
  \ScoreFunction{\TargetModel}(\Data{\SampleIdx}{\HiddenTwin{\SampleIdx}})}\}$
  \STATE \texttt{//Apply PPRM to obtain the confidence interval.}
  \STATE $[\ConfidenceIntervalLower{\ObservationTime}{\Confidence}, \ConfidenceIntervalUpper{\ObservationTime}{\Confidence}] \gets \PPRMartingale(\MeasurementSequence, \totalSamples, \Confidence)$
  \IF{$\ConfidenceIntervalLower{\ObservationTime}{\Confidence} \geq \sumThreshold$}
  \STATE $\DetectionResult \gets \mbox{True}$;
  \BREAK
  \ENDIF
  \ENDFOR
  \ENSURE \DetectionResult
\end{algorithmic}
\end{algorithm}

\section{Proof of Theorem~\ref{theorem:fdr}} \label{app:derivation}

\begin{proof}
We use $[\totalSamples]$ to represent $\{1,2,\dots,\totalSamples\}$.
We study the probability that under \nullHypothesis there exists a confidence interval whose lower bound is 
no smaller than a preselected threshold \sumThreshold ($\sumThreshold > \totalSamples / 2$), 
i.e., $\ConfidenceIntervalLower{\ObservationTime}{\Confidence} \geq \sumThreshold$. We have:
\begin{align*}
  \lefteqn{\cprob{\big}{\exists \ObservationTime \in [\totalSamples]:\ConfidenceIntervalLower{\ObservationTime}{\Confidence} \geq \sumThreshold}{\nullHypothesis}} \\
  & = \cprob{\big}{\exists \ObservationTime \in [\totalSamples]:\ConfidenceIntervalLower{\ObservationTime}{\Confidence} \geq \sumThreshold}{\TotalSuccess \geq \sumThreshold, \nullHypothesis} \times \cprob{\big}{\TotalSuccess \geq \sumThreshold}{\nullHypothesis} \\
  & \qquad + \cprob{\big}{\exists \ObservationTime \in [\totalSamples]:\ConfidenceIntervalLower{\ObservationTime}{\Confidence} \geq \sumThreshold}{\TotalSuccess < \sumThreshold , \nullHypothesis} \times \cprob{\big}{\TotalSuccess < \sumThreshold}{\nullHypothesis} \\
  & \leq  \cprob{\big}{\TotalSuccess \geq \sumThreshold}{\nullHypothesis}
   + \cprob{\big}{\exists \ObservationTime \in [\totalSamples]:\ConfidenceIntervalLower{\ObservationTime}{\Confidence} \geq \sumThreshold}{\TotalSuccess < \sumThreshold, \nullHypothesis} \\
  & = \cprob{\big}{\TotalSuccess \geq \sumThreshold}{\nullHypothesis}
    + \cprob{\big}{\exists \ObservationTime \in [\totalSamples]:\ConfidenceIntervalLower{\ObservationTime}{\Confidence} \geq \sumThreshold}{\TotalSuccess < \sumThreshold} .
\end{align*}
Here, for any $\sumThreshold > \totalSamples / 2$, according to a Chernoff bound~\cite[\eqnrefstatic{4.1}]{mitzenmacher2017:probability}, we have:
\begin{equation*}
  \cprob{\big}{\TotalSuccess \geq \sumThreshold}{\nullHypothesis} < \bigg(\frac{\exp\big(\frac{2\sumThreshold}{\totalSamples} - 1\big)}{\big(\frac{2\sumThreshold}{\totalSamples}\big)^{\frac{2\sumThreshold}{\totalSamples}}}\bigg)^{\frac{\totalSamples}{2}}.
\end{equation*}
Also, according to the guarantee of the confidence sequence~\cite{waudby2020:confidence}, we have:
\begin{equation*}
  \cprob{\big}{\exists \ObservationTime \in [\totalSamples]:\ConfidenceIntervalLower{\ObservationTime}{\Confidence} \geq \sumThreshold}{\TotalSuccess < \sumThreshold} \leq \Confidence.
\end{equation*}
As such, we have:
\begin{equation*}
  \cprob{\big}{\exists \ObservationTime \in [\totalSamples]: \ConfidenceIntervalLower{\ObservationTime}{\Confidence} \geq \sumThreshold}{\nullHypothesis} < \bigg(\frac{\exp\big(\frac{2\sumThreshold}{\totalSamples} - 1\big)}{\big(\frac{2\sumThreshold}{\totalSamples}\big)^{\frac{2\sumThreshold}{\totalSamples}}}\bigg)^{\frac{\totalSamples}{2}} + \Confidence.
\end{equation*}
Since our detection algorithm rejects \nullHypothesis if it finds a 
confidence interval whose lower bound satisfies $\ConfidenceIntervalLower{\ObservationTime}{\Confidence} \geq \sumThreshold$,
$\cprob{\big}{\exists \ObservationTime \in [\totalSamples]: \ConfidenceIntervalLower{\ObservationTime}{\Confidence} \geq \sumThreshold}{\nullHypothesis}$ 
is its false-detection probability. When we set a $\sumThreshold \in \{\ceil{\frac{\totalSamples}{2}}, \ldots,
\totalSamples\}$ and $\Confidence < \FalseDetectionRate$ such that
$\bigg(\frac{\exp\big(\frac{2\sumThreshold}{\totalSamples} -
  1\big)}{\big(\frac{2\sumThreshold}{\totalSamples}\big)^{\frac{2\sumThreshold}{\totalSamples}}}\bigg)^{\frac{\totalSamples}{2}}
\leq \FalseDetectionRate - \Confidence$, our detection algorithm has a
false-detection rate less than \FalseDetectionRate. In other words:
\begin{equation*}
  \cprob{\big}{\exists \ObservationTime \in \{1, 2, \dots, \totalSamples\}: \ConfidenceIntervalLower{\ObservationTime}{\Confidence} \geq \sumThreshold}{\nullHypothesis} < \FalseDetectionRate.
\end{equation*}
\end{proof}

\section{Score Functions Used in Image Classifier and Self-Supervised Visual Encoder}  \label{app:scorefunction}

\paragraph{Score function for image classifier}
We define the score function in
\algref{alg:score_function_image_classifier}.  Given an input image,
we first randomly generate \NumAugment perturbed versions, and then
obtain \NumAugment outputs using the perturbed images as inputs to the
target ML model.  We average the \NumAugment outputs and use the
negative (modified) entropy of the averaged output vector elements as the score.

\begin{algorithm}[!t]
  \caption{Score function for image classifier}  \label{alg:score_function_image_classifier}
  \begin{algorithmic}[1]
    \REQUIRE An input image \Data{\SampleIdx}{\genericInd},
    black-box access to the target ML model \TargetModel, number
    \NumAugment of image crops, number \NumClasses of classes, and
    the ground-truth label \GroundTruthLabel of
    \Data{\SampleIdx}{\genericInd}, if applicable;

    \STATE Generate \NumAugment perturbed images
    $\{\Data{\SampleIdx}{\genericInd}[\AugmentIdx]\}_{\AugmentIdx=1}^{\NumAugment}$
    by applying \NumAugment random, independent crops to
    \Data{\SampleIdx}{\genericInd};
    
  \FOR{$\AugmentIdx \gets 1, 2, \dots, \NumAugment$}
  \STATE $\ModelOutput{\AugmentIdx} \gets \TargetModel(\Data{\SampleIdx}{\genericInd}[\AugmentIdx])$;
  \ENDFOR
  \STATE $\ModelOutput \gets \frac{1}{\NumAugment} \sum_{\AugmentIdx=1}^{\NumAugment} \ModelOutput{\AugmentIdx}$;
  \IF{\GroundTruthLabel is known}
  \STATE $\Score{\SampleIdx}{\genericInd} \gets (1 - \arrComponent{\ModelOutput}{\GroundTruthLabel}) \log(\arrComponent{\ModelOutput}{\GroundTruthLabel}) 
      + \sum_{\ClassIdx \neq \GroundTruthLabel} \arrComponent{\ModelOutput}{\ClassIdx} \log(1-\arrComponent{\ModelOutput}{\ClassIdx})$
      where $\arrComponent{\ModelOutput}{\ClassIdx}$ denotes the \ClassIdx-th entry of \ModelOutput;
  \ELSE
  \STATE $\Score{\SampleIdx}{\genericInd} \gets \sum_{\ClassIdx=1}^{\NumClasses} \arrComponent{\ModelOutput}{\ClassIdx} \log(\arrComponent{\ModelOutput}{\ClassIdx})$;
  \ENDIF
  \ENSURE \Score{\SampleIdx}{\genericInd}
\end{algorithmic}
\end{algorithm}

\paragraph{Score function for self-supervised visual encoder}
We define the score function in
\algref{alg:score_function_visual_encoder}.  Given an input image, we
first randomly generate $\NumAugment$ of its perturbed versions (e.g.,
by random cropping and flipping), and then obtain $\NumAugment$
feature vectors using the perturbed images as inputs to the target
visual encoder.  Second, we compute the cosine similarity of every
pairs of feature vectors and return the sum of cosine similarities as
the score.

\begin{algorithm}[!t]
  \caption{Score function for visual encoder}  \label{alg:score_function_visual_encoder}
  \begin{algorithmic}[1]
  \REQUIRE An input image \Data{\SampleIdx}{\genericInd}, black-box
  access to the target visual encoder \TargetModel, and number of
  augmentations per image \NumAugment;
    
  \STATE Generate \NumAugment perturbed images
  $\{\Data{\SampleIdx}{\genericInd}[\AugmentIdx]\}_{\AugmentIdx=1}^{\NumAugment}$
  by applying \NumAugment random, independent crops and flips
  to \Data{\SampleIdx}{\genericInd};
  
  \STATE $\Score{\SampleIdx}{\genericInd} \gets 0$;
  \FOR{$\AugmentIdx \gets 1, 2, \dots, \NumAugment$}
  \FOR{$\AugmentIdxAlt \gets 1, 2, \dots, \NumAugment$}
  \STATE $\Score{\SampleIdx}{\genericInd} \gets \Score{\SampleIdx}{\genericInd} + \CosinSim(\TargetModel(\Data{\SampleIdx}{\genericInd}[\AugmentIdx]), \TargetModel(\Data{\SampleIdx}{\genericInd}[\AugmentIdxAlt]))$;
  \ENDFOR
  \ENDFOR
  \ENSURE \Score{\SampleIdx}{\genericInd};
\end{algorithmic}
\end{algorithm}

\section{Examples of Marked Images} \label{app:marked_examples}

\begin{figure*}[t]
    \centering
    \resizebox{\textwidth}{0.375\textwidth}{%
        \begin{tikzpicture}

            \definecolor{darkgray176}{RGB}{176,176,176}

            \begin{axis}[
                    hide axis,
                    tick align=outside,
                    tick pos=left,
                    x grid style={darkgray176},
                    xmin=-0.5, xmax=127.5,
                    xtick style={color=black},
                    y dir=reverse,
                    y grid style={darkgray176},
                    ymin=-0.5, ymax=63.5,
                    ytick style={color=black}
                ]
                \addplot graphics [includegraphics cmd=\pgfimage,xmin=-0.5, xmax=127.5, ymin=63.5, ymax=-0.5] {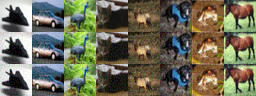};
            \end{axis}

        \end{tikzpicture}
    }
    \caption{Examples of marked CIFAR-10 images ($\MarkBound=10$). First row: raw images; Second row: published images; Last row: unpublished images.}
    \label{fig:cifar10_examples}
\end{figure*}

\begin{figure*}[t]
    \centering
    \resizebox{\textwidth}{0.375\textwidth}{%
        \begin{tikzpicture}

            \definecolor{darkgray176}{RGB}{176,176,176}

            \begin{axis}[
                    hide axis,
                    tick align=outside,
                    tick pos=left,
                    x grid style={darkgray176},
                    xmin=-0.5, xmax=127.5,
                    xtick style={color=black},
                    y dir=reverse,
                    y grid style={darkgray176},
                    ymin=-0.5, ymax=63.5,
                    ytick style={color=black}
                ]
                \addplot graphics [includegraphics cmd=\pgfimage,xmin=-0.5, xmax=127.5, ymin=63.5, ymax=-0.5] {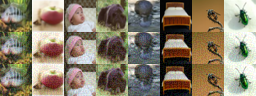};
            \end{axis}

        \end{tikzpicture}
    }
    \caption{Examples of marked CIFAR-100 images ($\MarkBound=10$). First row: raw images; Second row: published images; Last row: unpublished images.}
    \label{fig:cifar100_examples}
\end{figure*}

\begin{figure*}[t]
    \centering
    \resizebox{\textwidth}{0.375\textwidth}{%
        \begin{tikzpicture}

            \definecolor{darkgray176}{RGB}{176,176,176}

            \begin{axis}[
                    hide axis,
                    tick align=outside,
                    tick pos=left,
                    x grid style={darkgray176},
                    xmin=-0.5, xmax=127.5,
                    xtick style={color=black},
                    y dir=reverse,
                    y grid style={darkgray176},
                    ymin=-0.5, ymax=63.5,
                    ytick style={color=black}
                ]
                \addplot graphics [includegraphics cmd=\pgfimage,xmin=-0.5, xmax=127.5, ymin=63.5, ymax=-0.5] {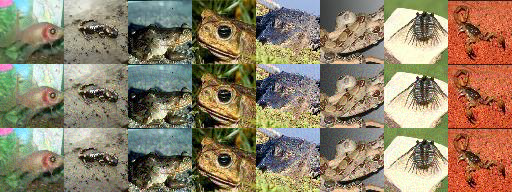};
            \end{axis}

        \end{tikzpicture}
    }
    \caption{Examples of marked TinyImageNet images ($\MarkBound=10$). First row: raw images; Second row: published images; Last row: unpublished images.}
    \label{fig:tinyimagenet_examples}
\end{figure*}

Some examples of marked data are shown in \figref{fig:cifar10_examples}, \figref{fig:cifar100_examples}, and \figref{fig:tinyimagenet_examples}.

\section{Introduction of Radioactive Data and UBW-C and Their Implementation Details} \label{app:baselines}

In \secref{sec:classifier:setup}, we considered two state-of-the-art
dataset auditing methods as baselines, namely Radioactive Data
(RData)~\cite{sablayrolles2020:radioactive} and Untargeted Backdoor
Watermark-Clean (UBW-C)~\cite{li2022:untargeted}, and applied them
into our problem settings.

\paragraph{RData}
Radioactive Data (RData) is a dataset auditing method including a marking
algorithm and a detection algorithm.  The marking algorithm adds
class-specific marks into a subset of the dataset.  The detection
algorithm tests a hypothesis on the parameters of the classifier layer
of a target ML model. Prior to the marking step, the data owner needs
to pretrain a feature extractor \PretrainedModel, referred as the
marking network, on the raw dataset; randomly samples a random unit
vector $\Carrier{\ClassIdx}$ for each class
$\ClassIdx=1,2,\dots,\NumClasses$; and randomly samples a subset of
her dataset to be marked. In the marking step, for an image
\Data{\SampleIdx} of class \ClassIdx in the sampled subset, the
marking algorithm adds a perturbation \Mark{\SampleIdx} to it such
that: the feature vector of $\Data{\SampleIdx} + \Mark{\SampleIdx}$
(i.e., $\PretrainedModel(\Data{\SampleIdx} + \Mark{\SampleIdx})$) has
a similar direction to $\Carrier{\ClassIdx}$; the marked image is
similar to its raw version in both feature space and pixel space; the
magnitude of the added mark is bounded, i.e.,
$\infinityNorm{\Mark{\SampleIdx}} \leq \MarkBound$.  Prior to the
detection step, the data owner needs to align the feature extractor of
the target classifier and the marking network by finding a linear
mapping between them. After this alignment, the detection algorithm
tests whether the cosine similarity between the parameter of the
classifier layer associated with class \ClassIdx and the mark
\Mark{\SampleIdx} follows an incomplete beta
distribution~\cite{iscen2017:memory} (the null hypothesis) or it is
large (the alternate hypothesis), and then combines the \NumClasses
tests to obtain a final result. If the combined p-value is smaller
than $0.05$, it concludes that the target classifier was trained on
the data owner's marked dataset; otherwise, it was not.

In the implementation of RData in our problem settings, we used their
open-source
code\footnote{\url{https://github.com/facebookresearch/radioactive_data}}.
We first trained a marking network on the raw dataset by our default
training method (introduced in ``training setting'' in
\secref{sec:classifier:setup}). Then, we followed our marking setting
(as introduced in \secref{sec:classifier:setup}) to constitute a
marked dataset. Specifically, we randomly sampled a subset of data
assumed to owned by a data owner and generated marked data for the raw
datum from the data owner by the marking algorithm of RData, using the
trained marking network and setting $\MarkBound=10$.  In the
application of the marking algorithm, if the labels assigned by the ML
practitioner are assumed to be known, we added a mark into the data
based on its label (i.e., we used the same mark for the data from the
same class but different marks for different classes); if the label
assignment is assumed to be unknown, we added one mark for all the
data (this method is ``RData (one mark)'' in
\secref{sec:classifier:setup}) or added marks based on ``coarse''
labels (that is ``RData (superclass)''). We used the superclass of
CIFAR-100 (\url{https://www.cs.toronto.edu/~kriz/cifar.html}) as the
``coarse'' labels in CIFAR-100 dataset.  For the TinyImageNet dataset,
we constructed ``coarse'' labels by querying ChatGPT
(\url{https://chat.openai.com/}); these ``coarse'' labels are shown in
\tblref{table:superclass_tiny}.  When given a target ML model, we
applied its detection algorithm that returned a combined p-value.  If
the combined p-value was smaller than $0.05$\footnote{Setting a
threshold of $0.05$ guarantees that the false-detection rate is
upper-bounded by $0.05$.}, we detected the use of marked data;
otherwise, we failed.

\begin{table*}[ht!]
  \centering
  \begin{tabular}{p{0.25\linewidth}p{0.7\linewidth}}
  \hline
  \textbf{Superclass} & \textbf{Class} \\
  \hline
  Aquatic Animals & European fire salamander, bullfrog, jellyfish, sea slug, spiny lobster, tailed frog, brain coral, goldfish, sea cucumber, American lobster \\
  Land Animals & koala, black widow, trilobite, scorpion, tarantula, centipede, boa constrictor, American alligator, dugong \\
  Birds & black stork, goose, king penguin, albatross \\
  Domestic Animals & Chihuahua, golden retriever, Persian cat, Yorkshire terrier, German shepherd, Egyptian cat, standard poodle, Labrador retriever, tabby \\
  Wild Animals & lion, chimpanzee, lesser panda, orangutan, brown bear, baboon, cougar, African elephant \\
  Insects and Arachnids & dragonfly, monarch, walking stick, grasshopper, ladybug, sulphur butterfly, fly, mantis, cockroach, bee \\
  Clothing and Accessories & academic gown, bikini, sandal, poncho, military uniform, cardigan, fur coat, miniskirt, swimming trunks, bow tie, kimono, vestment, sombrero, apron \\
  Transportation & trolleybus, gondola, lifeboat, jinrikisha, bullet train, convertible, school bus, police van, sports car, beach wagon, limousine, moving van \\
  Buildings and Structures & triumphal arch, cliff dwelling, butcher shop, fountain, steel arch bridge, barbershop, suspension bridge, barn, freight car, water tower, viaduct, dam, obelisk, beacon \\
  Household Items & beaker, snorkel, candle, Christmas stocking, dumbbell, turnstile, lawn mower, computer keyboard, parking meter, backpack, scoreboard, water jug, wok, dining table, pay-phone, sewing machine, hourglass, tractor, banister, pole, plate, sock, bathtub, torch, magnetic compass, spider web, frying pan, plunger, drumstick, birdhouse, gasmask, umbrella, stopwatch, rocking chair, teapot, sunglasses, flagpole, teddy, punching bag, beer bottle, lampshade, reel, refrigerator, rugby ball, pill bottle, broom, binoculars, space heater, chest, volleyball, iPod, bucket, maypole, desk, wooden spoon, syringe, remote control \\
  Food & pretzel, ice cream, cauliflower, ice lolly, meat loaf, espresso, potpie, mushroom, guacamole, bell pepper, pizza, orange, pomegranate, mashed potato, banana, lemon \\
  Natural Locations & coral reef, seashore, lakeside, alp, cliff \\
  Miscellaneous Objects & barrel, basketball, potter's wheel, Arabian camel, abacus, neck brace, oboe, projectile, confectionery, bighorn, chain, picket fence, hog, comic book, slug, guinea pig, nail, go-kart, ox, snail, gazelle, organ, altar, crane, pop bottle, bison \\
  Plants and Nature & acorn \\
  Technology and Electronics & cash machine, CD player \\
  Others & brass, thatch, cannon \\
  \hline
  \end{tabular}
  \vspace{1ex}
  \caption{Superclass of TinyImageNet generated by querying to ChatGPT.}
  \label{table:superclass_tiny}
  \end{table*}

\paragraph{UBW-C}
UBW-C is designed based on a clean-label untargeted backdoor
attack. It includes a marking (poisoning) algorithm that adds
perturbations to a subset of the dataset and a detection algorithm
that tests a hypothesis on the outputs of the target classifier.
Prior to the marking step, UBW-C also needs to train a surrogate model
on the raw dataset that is used to craft poisoned/marked data,
(strategically) selects a subset of data to be marked, and selects a
backdoor trigger. In the marking step, its marking algorithm crafts
marked data as $\Data{\SampleIdx} + \Mark{\SampleIdx}$
($\infinityNorm{\Mark{\SampleIdx}} \leq \MarkBound$) such that any
classifier trained on the marked data will have a desired backdoor
behavior that given a test image, the classifier outputs a much lower
confidence score of its ground-truth when the test image includes
the preselected backdoor trigger.  As such, the detection algorithm
conducts a pairwise t-test~\cite{larsen2005:introduction} to test if
the added backdoor trigger in the test images significantly reduces
their confidence scores of their ground-truths (by comparing with a
predetermined threshold \UBWCParameter).  If the returned
p-value\footnote{Such p-value is not the rigorous false-detection rate
of UWB-C.} is lower than $0.05$, it concludes that the classifier was
trained on the marked dataset; otherwise, it was not.

In the implementation of UWB-C in our problem settings, we used their
open-source
code\footnote{\url{https://github.com/THUYimingLi/Untargeted_Backdoor_Watermark}}.
We trained a surrogate model used for marking data and used the same
backdoor trigger as the previous work
(e.g.,~\cite{li2022:untargeted}).  Then, we followed our marking
setting to constitute a marked dataset. Different from the setting in
the previous work where the data owner has full control over the whole
dataset such that she can strategically select images to be marked
based on their gradient norms, our setting considers that only a
subset is from a data owner. We simulated by \emph{randomly} sampling a
subset of the dataset that was assumed from the data owner.  As such,
we applied its marking algorithm to generate marked data for the
sampled data, where we used $\MarkBound=16$ as suggested by the
previous work.  When given a target ML model, we applied its detection
algorithm to test if the model was trained on the marked dataset based
on the p-value.  In the implementation of the detection algorithm, we
used $\UBWCParameter=0.2$ and $\UBWCParameter=0.25$ as suggested by
the previous work (e.g.~\cite{li2022:untargeted}).

\section{Correlation Analysis} \label{app:correlation}

To evaluate the detection effectiveness of our method, one metric we
applied is the minimal percentage of published data in the training
set that resulted in detection (i.e., $\MinimalPercentage$).  However,
finding such $\MinimalPercentage$ is highly costly since we need to
test the detection on ML models trained on datasets containing
different percentages of published data under all the settings. We
found that $\frac{\NumberQueriedPublished}{\TrainDatasetSize}$ and
$\MinimalPercentage$ are strongly correlated by a correlation
analysis. We conducted this correlation analysis by collecting $240$
pairs of $\frac{\NumberQueriedPublished}{\TrainDatasetSize}$ and
$\MinimalPercentage$ values and calculating their Pearson correlation
coefficient~\cite{cohen2009:pearson}.  The resulted correlation
coefficient was $0.66$ (a correlation coefficient between $0.6$ and
$0.79$ is generally considered as a strong correlation) and the
p-value was $2.16\times 10^{-31}$.  So, to evaluate the effectiveness
of the detection method, we fixed the percentage of published data in
the training set (e.g., $10\%$ as the default) and used
$\frac{\NumberQueriedPublished}{\TrainDatasetSize}$ as a surrogate for
$\MinimalPercentage$.

\section{Impact of ML Model Architecture and Hyperparameters} \label{app:hyperparameters}

In \secref{sec:classifier:results}, we explored the impact of the ML
practitioner's model architecture and the data owner's hyperparameters
for our proposed method on detection, such as the utility bound
$\MarkBound$, the feature extractor $\PretrainedModel$ used to
generate marked data, the upper bound $\FalseDetectionRate$ on the
false-detection rate, and the number $\NumAugment$ of sampled
perturbations per image in detection.

\paragraph{Across different ML model architectures}
\tblref{table:classifier:different_architectures} shows the detection
results of our method on ML models across various architectures (e.g.,
VGG16~\cite{simonyan2015:very},
ConvNet64~\cite{krizhevsky2012:imagenet},
ConvNetBN~\cite{ioffe2015:batch}, and
MobileNetV2~\cite{sandler2018:mobilenetv2}) chosen by the ML
practitioner. The results demonstrate that our method is highly
effective across different ML model architectures, yielding a $20/20$
$\DetectionSuccessRate$ in all cases.

\begin{table*}[ht!]
    \centering
    \begin{tabular}{@{}lr@{\hspace{0.5em}}rr@{\hspace{0.5em}}rr@{\hspace{0.5em}}rr@{\hspace{0.5em}}rr@{\hspace{0.5em}}rr@{\hspace{0.5em}}rr@{\hspace{0.5em}}r@{}}
    \toprule
    \multicolumn{1}{c}{} &  \multicolumn{2}{c}{$\Knowledge{1}$}& \multicolumn{2}{c}{$\Knowledge{2}$}& \multicolumn{2}{c}{$\Knowledge{3}$}& \multicolumn{2}{c}{$\Knowledge{4}$}\\ 
    &  \multicolumn{1}{c}{$\DetectionSuccessRate$} & \multicolumn{1}{c}{$\frac{\NumberQueriedPublished}{\TrainDatasetSize}$} & \multicolumn{1}{c}{$\DetectionSuccessRate$} & \multicolumn{1}{c}{$\frac{\NumberQueriedPublished}{\TrainDatasetSize}$} & \multicolumn{1}{c}{$\DetectionSuccessRate$} & \multicolumn{1}{c}{$\frac{\NumberQueriedPublished}{\TrainDatasetSize}$} & \multicolumn{1}{c}{$\DetectionSuccessRate$} & \multicolumn{1}{c}{$\frac{\NumberQueriedPublished}{\TrainDatasetSize}$} \\ 
    \midrule 
    ResNet18 & $20/20$ & $0.19\%$ & $20/20$ & $0.20\%$ & $20/20$ & $0.59\%$ & $20/20$ & $0.60\%$ \\
    VGG16 & $20/20$ & $0.30\%$ & $20/20$ & $0.32\%$ & $20/20$ & $0.69\%$ & $20/20$ & $0.71\%$ \\ 
    ConvNet64 & $20/20$ & $0.72\%$ & $20/20$ & $2.98\%$ & $20/20$ & $0.90\%$ & $20/20$ & $3.89\%$ \\
    ConvNetBN & $20/20$ & $0.18\%$ & $20/20$ & $0.32\%$ & $20/20$ & $0.34\%$ & $20/20$ & $0.56\%$ \\ 
    MobileNetV2 & $20/20$ & $0.38\%$ & $20/20$ & $1.46\%$ & $20/20$ & $0.74\%$ & $20/20$ & $2.14\%$ \\
    \bottomrule 
    \end{tabular}
    \vspace{1ex}
    \caption{The performance of our method on detecting use of
      published data in ML model across different architectures,
      trained on CIFAR-100. All results are averaged over 20
      experiments.}
    \vspace{-4ex}
    \label{table:classifier:different_architectures}    
\end{table*}

\begin{figure*}[t]
    \centering

    \begin{subfigure}[b]{\textwidth}
        \centering
        \resizebox{\textwidth}{0.125\textwidth}{%
            \begin{tikzpicture}
                \definecolor{darkgray176}{RGB}{176,176,176}
                \begin{axis}[
                        hide axis,
                        tick align=outside,
                        tick pos=left,
                        x grid style={darkgray176},
                        xmin=-0.5, xmax=127.5,
                        xtick style={color=black},
                        y dir=reverse,
                        y grid style={darkgray176},
                        ymin=-0.5, ymax=63.5,
                        ytick style={color=black}
                    ]
                    \addplot graphics [includegraphics cmd=\pgfimage,xmin=-0.5, xmax=127.5, ymin=63.5, ymax=-0.5] {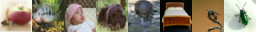};
                \end{axis}
            \end{tikzpicture}
        }
        \caption{Raw images}
        \label{fig:eps0}
    \end{subfigure}
    \vfill
    
    \begin{subfigure}[b]{\textwidth}
        \centering
        \resizebox{\textwidth}{0.125\textwidth}{%
            \begin{tikzpicture}
                \definecolor{darkgray176}{RGB}{176,176,176}
                \begin{axis}[
                        hide axis,
                        tick align=outside,
                        tick pos=left,
                        x grid style={darkgray176},
                        xmin=-0.5, xmax=127.5,
                        xtick style={color=black},
                        y dir=reverse,
                        y grid style={darkgray176},
                        ymin=-0.5, ymax=63.5,
                        ytick style={color=black}
                    ]
                    \addplot graphics [includegraphics cmd=\pgfimage,xmin=-0.5, xmax=127.5, ymin=63.5, ymax=-0.5] {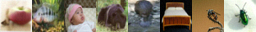};
                \end{axis}
            \end{tikzpicture}
        }
        \caption{$\MarkBound=2$}
        \label{fig:eps2}
    \end{subfigure}
    \vfill
    
    \begin{subfigure}[b]{\textwidth}
        \centering
        \resizebox{\textwidth}{0.125\textwidth}{%
            \begin{tikzpicture}
                \definecolor{darkgray176}{RGB}{176,176,176}
                \begin{axis}[
                        hide axis,
                        tick align=outside,
                        tick pos=left,
                        x grid style={darkgray176},
                        xmin=-0.5, xmax=127.5,
                        xtick style={color=black},
                        y dir=reverse,
                        y grid style={darkgray176},
                        ymin=-0.5, ymax=63.5,
                        ytick style={color=black}
                    ]
                    \addplot graphics [includegraphics cmd=\pgfimage,xmin=-0.5, xmax=127.5, ymin=63.5, ymax=-0.5] {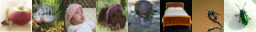};
                \end{axis}
            \end{tikzpicture}
        }
        \caption{$\MarkBound=4$}
        \label{fig:eps4}
    \end{subfigure}
    \vfill
    
    \begin{subfigure}[b]{\textwidth}
        \centering
        \resizebox{\textwidth}{0.125\textwidth}{%
            \begin{tikzpicture}
                \definecolor{darkgray176}{RGB}{176,176,176}
                \begin{axis}[
                        hide axis,
                        tick align=outside,
                        tick pos=left,
                        x grid style={darkgray176},
                        xmin=-0.5, xmax=127.5,
                        xtick style={color=black},
                        y dir=reverse,
                        y grid style={darkgray176},
                        ymin=-0.5, ymax=63.5,
                        ytick style={color=black}
                    ]
                    \addplot graphics [includegraphics cmd=\pgfimage,xmin=-0.5, xmax=127.5, ymin=63.5, ymax=-0.5] {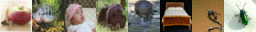};
                \end{axis}
            \end{tikzpicture}
        }
        \caption{$\MarkBound=6$}
        \label{fig:eps6}
    \end{subfigure}
    \vfill
    
    \begin{subfigure}[b]{\textwidth}
        \centering
        \resizebox{\textwidth}{0.125\textwidth}{%
            \begin{tikzpicture}
                \definecolor{darkgray176}{RGB}{176,176,176}
                \begin{axis}[
                        hide axis,
                        tick align=outside,
                        tick pos=left,
                        x grid style={darkgray176},
                        xmin=-0.5, xmax=127.5,
                        xtick style={color=black},
                        y dir=reverse,
                        y grid style={darkgray176},
                        ymin=-0.5, ymax=63.5,
                        ytick style={color=black}
                    ]
                    \addplot graphics [includegraphics cmd=\pgfimage,xmin=-0.5, xmax=127.5, ymin=63.5, ymax=-0.5] {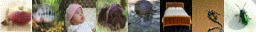};
                \end{axis}
            \end{tikzpicture}
        }
        \caption{$\MarkBound=8$}
        \label{fig:eps8}
    \end{subfigure}
    \vfill
    
    \begin{subfigure}[b]{\textwidth}
        \centering
        \resizebox{\textwidth}{0.125\textwidth}{%
            \begin{tikzpicture}
                \definecolor{darkgray176}{RGB}{176,176,176}
                \begin{axis}[
                        hide axis,
                        tick align=outside,
                        tick pos=left,
                        x grid style={darkgray176},
                        xmin=-0.5, xmax=127.5,
                        xtick style={color=black},
                        y dir=reverse,
                        y grid style={darkgray176},
                        ymin=-0.5, ymax=63.5,
                        ytick style={color=black}
                    ]
                    \addplot graphics [includegraphics cmd=\pgfimage,xmin=-0.5, xmax=127.5, ymin=63.5, ymax=-0.5] {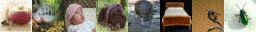};
                \end{axis}
            \end{tikzpicture}
        }
        \caption{$\MarkBound=10$}
        \label{fig:eps10}
    \end{subfigure}

    \caption{Examples of marked CIFAR-100 images under different $\MarkBound$.}
    \label{fig:different_eps}
\end{figure*}

\paragraph{Impact of $\MarkBound$}
\tblref{table:classifier:markbound} shows the impact of $\MarkBound$
on the effectiveness of our method, demonstrating a trade-off between
detection effectiveness/efficiency and data utility measured by
$\Accuracy$ of a ML model trained on it.  $\MarkBound$ controls the
upper bound of utility difference between the published data and raw
data. For images, it also controls the imperceptibility of the added
marks in the published data. A smaller $\MarkBound$ yields less
perceptible marks (see some examples in \figref{fig:different_eps})
and allows the published data to preserve more utility of the raw
data, i.e., the trained ML model has a higher $\Accuracy$. However,
the choice of a smaller $\MarkBound$ will reduce the distinction
between the published data and unpublished data, and thus it slightly
degrades the detection effectiveness/efficiency of our method,
yielding a lower $\DetectionSuccessRate$ (when $\MarkBound = 2$) and a
higher $\frac{\NumberQueriedPublished}{\TrainDatasetSize}$.

\begin{table*}[ht!]
    \centering
    \begin{tabular}{@{}lr@{\hspace{0.5em}}rr@{\hspace{0.5em}}rr@{\hspace{0.5em}}rr@{\hspace{0.5em}}rr@{\hspace{0.5em}}rr@{\hspace{0.5em}}rr@{\hspace{0.5em}}r@{}}
    \toprule
    \multicolumn{1}{c}{} & \multicolumn{1}{c}{\multirow{2}{*}{$\Accuracy \%$}}& \multicolumn{2}{c}{$\Knowledge{1}$}& \multicolumn{2}{c}{$\Knowledge{2}$}& \multicolumn{2}{c}{$\Knowledge{3}$}& \multicolumn{2}{c}{$\Knowledge{4}$}\\ 
    & & \multicolumn{1}{c}{$\DetectionSuccessRate$} & \multicolumn{1}{c}{$\frac{\NumberQueriedPublished}{\TrainDatasetSize}$} & \multicolumn{1}{c}{$\DetectionSuccessRate$} & \multicolumn{1}{c}{$\frac{\NumberQueriedPublished}{\TrainDatasetSize}$} & \multicolumn{1}{c}{$\DetectionSuccessRate$} & \multicolumn{1}{c}{$\frac{\NumberQueriedPublished}{\TrainDatasetSize}$} & \multicolumn{1}{c}{$\DetectionSuccessRate$} & \multicolumn{1}{c}{$\frac{\NumberQueriedPublished}{\TrainDatasetSize}$}\\ 
    \midrule 
    $\MarkBound = 2$ & $74.99$ & $20/20$ & $2.69\%$ & $20/20$ & $2.87\%$ & $11/20$ & $9.48\%$ & $12/20$ & $9.77\%$ \\
    $\MarkBound = 4$ & $74.84$ & $20/20$ & $0.80\%$ & $20/20$ & $1.00\%$ & $20/20$ & $4.25\%$ & $20/20$ & $4.38\%$ \\ 
    $\MarkBound = 6$ & $74.64$ & $20/20$ & $0.43\%$ & $20/20$ & $0.50\%$ &  $20/20$ & $1.54\%$ & $20/20$ & $1.97\%$ \\  
    $\MarkBound = 8$ & $74.39$ & $20/20$ & $0.25\%$ & $20/20$ & $0.28\%$ &  $20/20$ & $0.85\%$ & $20/20$ & $1.08\%$ \\ 
    $\MarkBound = 10$ & $74.29$ & $20/20$ & $0.19\%$ & $20/20$ & $0.20\%$ & $20/20$ & $0.59\%$ & $20/20$ & $0.60\%$ \\
    \bottomrule 
    \end{tabular}
    \vspace{1ex}
    \caption{The impact of $\MarkBound$ on the performance of our
      method applied on CIFAR-100.  $\MarkBound = 10$ is
      the default. All results are averaged over 20 experiments.}
    \vspace{-4ex}
    \label{table:classifier:markbound}    
\end{table*}

\paragraph{Impact of feature extractor \PretrainedModel used in
  marked data generation} \tblref{table:classifier:marking_network}
shows the impact of feature extractor $\PretrainedModel$ used in
the marking algorithm of our method when applied to images.  Using
ResNet18~\cite{he2016:deep} pretrained on ImageNet,
VGG16~\cite{simonyan2015:very} pretrained on ImageNet, and
CLIP~\cite{radford2021:learning} to generate marked data yielded
similar detection performance. This demonstrates that our method is
highly effective and efficient when a good feature extractor (e.g.,
pretrained on a large-scale image dataset) is used in
the marking algorithm. More surprisingly, when an untrained/randomly
initialized feature extractor (e.g., untrained ResNet18) was used, the
efficiency of our method degraded slightly, yielding a slightly higher
$\frac{\NumberQueriedPublished}{\TrainDatasetSize}$, but it
still achieved a high $\DetectionSuccessRate$ of $20/20$.  When there
is no feature extractor available to the data owner or she does not
have computing resources to craft the marked data using a feature
extractor, she can craft $\Mark{\SampleIdx}$ by randomly sampling from
a Gaussian distribution and clipping it such that
$\infinityNorm{\Mark{\SampleIdx}} \leq \MarkBound$, or by
randomly sampling from a Bernoulli distribution
(i.e., each pixel of a mark is independently and randomly chosen to be either
$\MarkBound$ or $-\MarkBound$ with equal probability). These variants of
our method, denoted by ``Gaussian mark'' and ``Bernoulli mark'' 
respectively in \tblref{table:classifier:marking_network}, were still effective to
detect the use of data but achieved a higher
$\frac{\NumberQueriedPublished}{\TrainDatasetSize}$, compared
with the methods where a feature extractor was used.

\begin{table*}[ht!]
    \centering
    \begin{tabular}{@{}lr@{\hspace{0.8em}}rr@{\hspace{0.8em}}rr@{\hspace{0.8em}}rr@{\hspace{0.8em}}rr@{\hspace{0.8em}}rr@{\hspace{0.8em}}rr@{\hspace{0.8em}}r@{}}
    \toprule
    \multicolumn{1}{c}{} & \multicolumn{2}{c}{$\Knowledge{1}$}& \multicolumn{2}{c}{$\Knowledge{2}$}& \multicolumn{2}{c}{$\Knowledge{3}$}& \multicolumn{2}{c}{$\Knowledge{4}$}\\ 
    & \multicolumn{1}{c}{$\DetectionSuccessRate$} & \multicolumn{1}{c}{$\frac{\NumberQueriedPublished}{\TrainDatasetSize}$} & \multicolumn{1}{c}{$\DetectionSuccessRate$} & \multicolumn{1}{c}{$\frac{\NumberQueriedPublished}{\TrainDatasetSize}$} & \multicolumn{1}{c}{$\DetectionSuccessRate$} & \multicolumn{1}{c}{$\frac{\NumberQueriedPublished}{\TrainDatasetSize}$} & \multicolumn{1}{c}{$\DetectionSuccessRate$} & \multicolumn{1}{c}{$\frac{\NumberQueriedPublished}{\TrainDatasetSize}$} \\ 
    \midrule 
    ResNet18 & $20/20$ & $0.19\%$ & $20/20$ & $0.20\%$ & $20/20$ & $0.59\%$ & $20/20$ & $0.60\%$ \\
    VGG16 & $20/20$ & $0.20\%$ & $20/20$ & $0.22\%$ & $20/20$ & $0.51\%$ & $20/20$ & $0.72\%$ \\
    CLIP & $20/20$ & $0.21\%$ & $20/20$ & $0.26\%$ & $20/20$ & $0.65\%$ & $20/20$ & $0.66\%$ \\
    Untrained & $20/20$ & $0.28\%$ & $20/20$ & $0.32\%$ & $20/20$ & $1.14\%$ & $20/20$ & $1.26\%$ \\
    Gaussian mark & $20/20$ & $0.54\%$ & $20/20$ & $0.56\%$ & $20/20$ & $2.68\%$ & $20/20$ & $2.86\%$ \\
    Bernoulli mark & $20/20$ & $0.27\%$ & $20/20$ & $0.31\%$ & $20/20$ & $1.24\%$ & $20/20$ & $2.08\%$ \\   
    \bottomrule 
    \end{tabular}
    \vspace{1ex}
    \caption{The impact of the choice of feature extractor used in
      marked-data generation on the performance of our method applied
      on CIFAR-100.  ResNet18 and VGG16 were pretrained on ImageNet
      while CLIP was developed by OpenAI, pretrained on millions
      image-text pairs from the Internet.  ``Untrained'' refers to
      using a randomly initialized ResNet18. ``Gaussian mark'' and ``Bernoulli mark''
      refer to the settings where there is no feature extractor
      available.  ResNet18 (pretrained on ImageNet) was the default in
      other experiments.  All results were averaged over 20
      experiments.}
    \vspace{-4ex}
    \label{table:classifier:marking_network}    
\end{table*}

\paragraph{Different choices of upper bound $\FalseDetectionRate$ on false-detection rate}
\tblref{table:classifier:different_p} shows the performance of our
data auditing method under different upper bounds on the
false-detection rate, i.e., $\FalseDetectionRate$.  Under all choices
of $\FalseDetectionRate$, our method achieved a
$\DetectionSuccessRate$ of $20/20$. When we set a smaller
$\FalseDetectionRate$, we can control the false-detection rate to a
smaller level but require more queries to detect the use of the
published data in model training.

\begin{table*}[ht!]
  \centering
  \begin{tabular}{@{}lr@{\hspace{0.5em}}rr@{\hspace{0.5em}}rr@{\hspace{0.5em}}rr@{\hspace{0.5em}}rr@{\hspace{0.5em}}rr@{\hspace{0.5em}}rr@{\hspace{0.5em}}r@{}}
  \toprule
  \multicolumn{1}{c}{} & \multicolumn{2}{c}{$\Knowledge{1}$}& \multicolumn{2}{c}{$\Knowledge{2}$}& \multicolumn{2}{c}{$\Knowledge{3}$}& \multicolumn{2}{c}{$\Knowledge{4}$}\\
  & \multicolumn{1}{c}{$\DetectionSuccessRate$} & \multicolumn{1}{c}{$\frac{\NumberQueriedPublished}{\TrainDatasetSize}$} & \multicolumn{1}{c}{$\DetectionSuccessRate$} & \multicolumn{1}{c}{$\frac{\NumberQueriedPublished}{\TrainDatasetSize}$} & \multicolumn{1}{c}{$\DetectionSuccessRate$} & \multicolumn{1}{c}{$\frac{\NumberQueriedPublished}{\TrainDatasetSize}$} & \multicolumn{1}{c}{$\DetectionSuccessRate$} & \multicolumn{1}{c}{$\frac{\NumberQueriedPublished}{\TrainDatasetSize}$} \\ 
  \midrule 
   $\FalseDetectionRate=0.05$ & $20/20$ & $0.19\%$ & $20/20$ & $0.20\%$ & $20/20$ & $0.59\%$ & $20/20$ & $0.60\%$  \\ 
   $\FalseDetectionRate=0.01$ & $20/20$ & $0.32\%$ & $20/20$ & $0.36\%$ & $20/20$ & $0.89\%$ & $20/20$ & $0.78\%$ \\ 
   $\FalseDetectionRate=10^{-3}$ & $20/20$ & $0.39\%$ & $20/20$ & $0.42\%$ & $20/20$ & $1.33\%$ & $20/20$ & $1.21\%$ \\ 
   $\FalseDetectionRate=10^{-4}$ & $20/20$ & $0.61\%$ & $20/20$ & $0.70\%$ & $20/20$ & $1.70\%$ & $20/20$ & $1.86\%$ \\ 
   $\FalseDetectionRate=10^{-5}$ & $20/20$ & $0.79\%$ & $20/20$ & $0.86\%$ & $20/20$ & $1.90\%$ & $20/20$ & $2.39\%$ \\ 
   $\FalseDetectionRate=10^{-6}$ & $20/20$ & $0.89\%$ & $20/20$ & $0.97\%$ & $20/20$ & $2.96\%$ & $20/20$ & $2.43\%$ \\
  \bottomrule 
  \end{tabular} 
  \vspace{1ex}
  \caption{CIFAR-100 auditing results under different upper bounds of false-detection rate ($0.05$ is the default).
  We set $\Confidence = \FalseDetectionRate / 2$.
  All results are averaged over 20 experiments.}
  \vspace{-4ex}
  \label{table:classifier:different_p}
  \end{table*}

\paragraph{Impact of $\NumAugment$} \tblref{table:classifier:numaug}
shows the impact of $\NumAugment$ on the detection performance of our
method.  As shown in \tblref{table:classifier:numaug}, increasing
$\NumAugment$ led to a smaller
$\frac{\NumberQueriedPublished}{\TrainDatasetSize}$ but a larger
$\NumQuery$ (since $\NumQuery = 2 \times \NumAugment \times
\NumberQueriedPublished$).  Therefore, setting a larger $\NumAugment$
improves the detection efficiency but degrades the cost efficiency of
our proposed method.

\begin{table*}[ht!]
    \centering
      \begin{tabular}{@{}lr@{\hspace{0.8em}}rr@{\hspace{0.8em}}rr@{\hspace{0.8em}}rr@{\hspace{0.8em}}rr@{\hspace{0.8em}}rr@{\hspace{0.8em}}r@{}}
        \toprule
        \multicolumn{1}{c}{}& \multicolumn{3}{c}{$\Knowledge{1}$}& \multicolumn{3}{c}{$\Knowledge{2}$}& \multicolumn{3}{c}{$\Knowledge{3}$}& \multicolumn{3}{c}{$\Knowledge{4}$}\\ 
        & \multicolumn{1}{c}{$\DetectionSuccessRate$} & \multicolumn{1}{c}{$\frac{\NumberQueriedPublished}{\TrainDatasetSize}$} & \multicolumn{1}{c}{\NumQuery}& \multicolumn{1}{c}{$\DetectionSuccessRate$} & \multicolumn{1}{c}{$\frac{\NumberQueriedPublished}{\TrainDatasetSize}$} & \multicolumn{1}{c}{\NumQuery}& \multicolumn{1}{c}{$\DetectionSuccessRate$} & \multicolumn{1}{c}{$\frac{\NumberQueriedPublished}{\TrainDatasetSize}$} & \multicolumn{1}{c}{\NumQuery}& \multicolumn{1}{c}{$\DetectionSuccessRate$} & \multicolumn{1}{c}{$\frac{\NumberQueriedPublished}{\TrainDatasetSize}$} & \multicolumn{1}{c}{\NumQuery}\\ 
        \midrule 
        $\NumAugment=1$ & $20/20$ & $0.41\%$ & 410 & $20/20$ & $0.50\%$ & 495 & $20/20$ & $6.71\%$ & 6{,}702 & $0/20$ & $10.00\%$ & 10{,}000 \\ 
        $\NumAugment=2$ & $20/20$ & $0.36\%$ & 709 & $20/20$ & $0.41\%$ & 814 & $20/20$ & $3.21\%$ & 6{,}408 & $20/20$ & $6.40\%$ & 12{,}796 \\ 
        $\NumAugment=4$ & $20/20$ & $0.34\%$ & 1{,}334 & $20/20$ & $0.37\%$ & 1{,}457 & $20/20$ & $1.53\%$ & 6{,}107 & $20/20$ & $2.77\%$ & 11{,}047 \\ 
        $\NumAugment=8$ & $20/20$ & $0.26\%$ & 2{,}038 & $20/20$ & $0.31\%$ & 2{,}419 & $20/20$ & $0.82\%$ & 6{,}509 & $20/20$ & $1.08\%$ & 8{,}632 \\ 
        $\NumAugment=16$ & $20/20$ & $0.19\%$ & 2{,}964 & $20/20$ & $0.20\%$ & 3{,}087 & $20/20$ & $0.59\%$ & 9{,}312 & $20/20$ & $0.60\%$ & 9{,}492 \\ 
        $\NumAugment=32$ & $20/20$ & $0.30\%$ & 9{,}364 & $20/20$ & $0.32\%$ & 9{,}975 & $20/20$ & $0.51\%$ & 16{,}192 & $20/20$ & $0.48\%$ & 15{,}146 \\
        \bottomrule 
        \end{tabular}
    \vspace{1ex}
    \caption{The impact of $\NumAugment$ on the performance of our
      method applied on CIFAR-100.  $\NumAugment = 16$ was the default
      in other experiments. All results were averaged over 20
      experiments.}
    \vspace{-4ex}
    \label{table:classifier:numaug}    
\end{table*}

\end{document}